\documentclass[11pt]{article}
\usepackage[T1]{fontenc}
\usepackage{lmodern,amsmath,amssymb,booktabs,longtable,array,graphicx,microtype,placeins}
\usepackage[margin=0.8in]{geometry}
\usepackage[nobottomtitles*]{titlesec}

\usepackage[hidelinks]{hyperref}
\title{Retrospective Analysis of Team-Following Trading Profitability in Pump.fun Meme Coins Using Funding Networks}
\author{Fail Gafarov\\ Institute of Computational Mathematics and Information Technologies \\
Kazan Federal University \\
Kazan, Russia \\
fail.gafarov@kpfu.ru, fgafarov@yandex.ru }
\date{8 November 2026}
\begin{document}
\maketitle
\begin{abstract}
If the right token-creation teams had been known, how much could following them have returned, and what is left when only information available before each launch is used? We study ten separately recorded Pump.fun windows (April 2024 to September 2026), each reconstructed independently from real-SOL reserve states and finite-capital accounts that start with one SOL. Teams are directed funding groups ranked in hindsight. When every team exits at its own best reserve threshold, chosen in hindsight, the main account ends above one SOL in all 8 windows that have a qualifying team (1.10 to 11.32 SOL), but free cash exceeds one SOL in only 5, and marks of positions that never reached their threshold make up 17--94\% of equity. If teams are recognised online, from tokens whose first hour has completed, 9 of 19 teams become identifiable only after their last launch; the remaining 10 end above one SOL in 10 cases, with free cash above one SOL in 7; funding links in this step still come from the full window, so even these results are not a fully realisable signal. A large model balance, a result above random cohorts that repeat the threshold search, and a statistical rejection in the selection-matched nulls (3 of 40 tests after Holm adjustment) are three different statements, and these nulls condition on the discovered funding graph, so they do not calibrate the whole selection procedure. Midpoint tests describe a further source of apparent opportunity. A detailed analysis of the best and the worst window, with every analysis parameter varied, shows that the worst result, which is close to break-even, rises several-fold with a stop rule or another funding-graph construction. The results are conditional retrospective model accounts, not executable profits, and not evidence of a causal effect of the SOL price.
\end{abstract}
\section{Question and scope}
The appeal of a successful meme-coin backtest is immediate: identify creators that repeatedly launch and pump tokens, follow their subsequent launches and sell before dump. Each step has an information requirement. A group identified at the end of a window may not have been recognisable at its beginning, and a large marked balance may never have supplied reusable cash. We ask how such apparent opportunities differ across separately recorded Pump.fun windows, and how those differences change when selection seeds, finite capital, observation quality and hindsight search are made explicit.

Every funding graph, group catalogue, ranking, random control and temporal test is reconstructed within its own window. No identity learned in one period qualifies a group in another. Portfolio values are denominated in SOL with the same initial capital. 
A team in this paper is a candidate coordination group defined by directed funding links, not a verified organisation, beneficial owner or fraud label. This study concerns conditional following returns, not classification of rug pulls. The primary trading calculation uses recorded SOL reserve states and never recorded token prices or quantities. The boundary of the admitted historical and counterfactual reserve is strictly 85 real SOL. Results are hypothetical finite-size accounts under a declared reference mechanism, rather than reconstructed fills. 

\section{Related research}

This narrative review connects the research question to evidence on market manipulation, entity inference, decentralized-market execution and strategy evaluation. Sources were selected for their relevance to these methodological decisions, rather than as an exhaustive systematic review. We distinguish published studies from the explicitly versioned recent preprints. Findings from another chain, exchange or sampling frame motivate hypotheses; they are not prevalence estimates or fraud labels for the present research.

\subsection{Cryptocurrency manipulation and the outsider's information disadvantage}
Gandal et al. connect suspicious activity on Mt.\ Gox with Bitcoin price movements~\cite{gandal}. Their exchange-specific transaction evidence illustrates why an observed price increase need not represent independent demand. For the present study, the implication is methodological: a token's maximum recorded appreciation is an outcome to explain, not a sufficient definition of a successful opportunity for an outside trader. Kamps and Kleinberg develop price- and volume-based criteria for locating cryptocurrency pump-and-dump patterns, and compare anomaly flags with externally observed organized events~\cite{kamps}. This distinction matters because a spike-and-reversal pattern can flag an episode without identifying its organizer. Xu and Livshits instead investigate publicly coordinated pump events and evaluate a predictive trading approach~\cite{xu}. Their work is particularly close to the motivating trading question: information available before an announcement differs economically from a profitable token chosen after its peak. Their exchange-based results cannot establish achievable returns for newly launched bonding-curve tokens.

Li, Shin and Wang document short-lived changes in prices, trading activity and volatility around organized pumps and examine evidence consistent with transfers from outsiders to insiders~\cite{li_pump}. Their wallet-level analysis and exchange-policy comparisons go beyond selecting large price multiples. Taken together, these studies motivate separating the detection of a suspicious event, the identification of related actors and the performance of a rule available to an outsider. Dhawan and \mbox{Putni\c{n}\v{s}} study pump-and-dump manipulation of cryptocurrencies and ask why traders take part in such schemes~\cite{dhawan}. Hamrick et al. follow pumps coordinated in Telegram and Discord channels and report modest short-term price gains that shrank over their sample~\cite{hamrick}. La Morgia et al. analyse and detect pump-and-dump and crowd-pump episodes organised by online communities~\cite{lamorgia}. 

\subsection{Wash trading, scam tokens and recent Solana evidence}
Victor and Weintraud identify self-trades and more complex wash-trading structures on the Ethereum order-book exchanges IDEX and EtherDelta~\cite{wash}. Their transaction-level approach shows why trading activity and economically independent participation need not coincide. It also supplies a useful boundary: the directed funding links used here are not a wash-trading detector. High activity, reciprocal funding or a shared ancestor is insufficient to transfer their wash-trade classification to our observations.

Xia et al. combine association heuristics and machine learning to characterize scam tokens and colluding addresses on Uniswap~\cite{xia}. Their analysis includes contract-level mechanisms as well as related actors. Werner et al. distinguish technical from economic security in their systematization of decentralized finance~\cite{werner}. This distinction prevents treating a price collapse, a vulnerable contract and a coordinated trading campaign as interchangeable events. The present article studies counterfactual following returns; it does not estimate the incidence of all these mechanisms or construct a supervised fraud classifier.

The closest platform-specific reference is Szwajcok et al.'s Pump.fun study, which distinguishes wash trading, creator-address obfuscation, coordinated selling and other manipulation mechanisms~\cite{factories}. Its attention to automated actors motivates creator and funding-link analysis. Its separate mechanism definitions also limit our interpretation: a funding-connected group need not satisfy any particular manipulation criterion.  Two recent Solana preprints sharpen the distinction between early warning and following returns. Chen et al. develop a behavior-guided identification and validation pipeline for Solana rug pulls, separating authority abuse, liquidity withdrawal and pump-and-dump patterns across several trading venues~\cite{chen_sol}. Li et al. use early trading observations to predict fraudulent memecoin outcomes and examine generalization between PumpFun and Raydium~\cite{li_sol}. These tasks have explicit outcome definitions and observation horizons. A classifier evaluated after an early observation interval is not a signal available immediately at token creation; classification performance is also not a self-financing portfolio return. 

Outside Pump.fun, Cernera et al. describe token spammers, one-day rug pulls and sniper bots on Ethereum and the BNB Smart Chain~\cite{cernera}. Mongardini and Mei study 34,988 meme coins on Ethereum, BNB Smart Chain, Solana and Base and find that many high-return tokens show signs of artificial growth, such as wash trading or liquidity-pool price inflation, often followed by profit extraction~\cite{mongardini}. Cong et al. detect wash trading on centralised exchanges from statistical regularities of the trade data~\cite{cong}. These works classify tokens or trading by mechanism.

\subsection{Wallet clustering and the interpretation of a team}\label{sec:wallet_literature}
Meiklejohn et al. demonstrate how transaction structure and address-clustering heuristics can reduce the anonymity of Bitcoin payments~\cite{meiklejohn}. The broader lesson is that an address is not necessarily an economic entity. Bitcoin setting has transaction structures that differ from Solana account-based chains. Victor proposes Ethereum-specific heuristics based on deposit-address reuse, airdrop participation and token authorization patterns~\cite{victor}. Their dependence on observed usage patterns is directly relevant to interpretation: a connection has evidential value only under the assumptions of the particular heuristic. 

This literature motivates two safeguards in our design. First, the word \emph{team} denotes a candidate coordination group, not a verified person, business or criminal organization. A common service, distributor or intermediary can connect unrelated users. Excluding highly active endpoints and varying graph parameters reduce dependence on particular graph choices, but cannot measure an identity error rate without external ground truth. Second, information must be timestamped. A graph reconstructed after a window ends may reveal connections unavailable before a launch. Freezing the midpoint graph and separately examining causally expanding funding knowledge distinguish retrospective association from a potentially observable signal.

\subsection{AMM mechanics, market impact and transaction ordering}
Angeris et al. analyze constant-product markets and their relation to reference prices under stated assumptions~\cite{angeris}. This supplies the mathematical reason to distinguish marginal price from the finite proceeds of liquidating a position. It does not establish the historical reserve units or configuration of a different protocol. We therefore use official Pump.fun documentation only to define a declared curve reference~\cite{pump,fees}, and test recorded rows against that reference rather than treating generic AMM theory as historical data verification.

Daian et al. document competition among arbitrage bots and the importance of transaction ordering on decentralized exchanges~\cite{daian}. Their Ethereum setting is not a direct measurement of the Solana execution process, but it establishes a relevant mechanism: observing a trade price does not guarantee obtaining that price after a signal. In our interpretation, execution latency, reordering and competition remain omitted determinants of realized trading performance. A deterministic order for reproducible data processing solves an accounting problem; it does not reconstruct the order in which an outsider's transaction would have executed.

These considerations connect directly to the finite position caps, conditional constant-product valuation, state-age restrictions and fee/slippage scenarios used below. They also explain why marking an unclosed position at the last admissible quote cannot be counted as reusable cash. Neither the constant-product reference nor the added execution costs constitutes an exact counterfactual replay of a shared historical pool. Profitability claims therefore remain conditional even when a ranking index is large and every cash debit and credit is arithmetically correct.

Lehar and Parlour compare the Uniswap automated market maker with a centralised order book, using all interactions up to December 2022, and report stable liquidity pools and no long-lived arbitrage opportunities~\cite{lehar}. This is evidence about a constant-product venue with liquidity providers; the Pump.fun curve is a protocol-owned pool, so we use the result only to motivate modelling the price impact of one's own position. Qin et al. quantify extractable value from sandwich attacks, arbitrage and liquidations on Ethereum~\cite{qin}. Together with Daian et al. they support the caution that an observed price is not an executable price a few seconds later; the entry-delay sensitivities of Sections~\ref{sec:deep} and~\ref{sec:strategies} are therefore descriptive and do not measure latency.

\subsection{Strategy search, random controls and temporal evaluation}
Sullivan et al. evaluate technical trading rules while accounting for the larger rule universe from which a reported winner is selected~\cite{sullivan}. White's Reality Check formalizes testing a best discovered model against a benchmark~\cite{white}, while Hansen's superior-predictive-ability test addresses sensitivity to irrelevant alternatives and the construction of the comparison statistic~\cite{hansen}. These works explain why beating a typical random trader is a different claim from beating the best result produced by a comparably extensive search. 
Bailey et al. examine backtest overfitting and propose a framework based on the relation between selected in-sample and subsequent performance~\cite{bailey}. Harvey, Liu and Zhu discuss the consequences of extensive multiple testing for financial discoveries~\cite{harvey}. We use these papers to motivate disclosure of the searched team and exit-policy universe, parameter surfaces and information cutoffs. 

We do not implement their full testing frameworks, estimate the probability of backtest overfitting, or import a numerical significance hurdle from a different asset-pricing problem. Our controls consequently have separate roles. The tokens-by-timing comparison describes how a selected cohort differs from uniform token choice and random timing under common cash rules. The selection-matched relabelling control repeats the ranking search while preserving group sizes, overlaps and selected token strata. The midpoint exercise asks whether groups identified earlier cover later launches in the same period. None substitutes for all the others: a conditional graph-label randomization does not rerun group discovery, and an out-of-sample segment chosen after inspecting the data is weaker evidence than a predeclared future evaluation. Our additional causal-expansion exercise is therefore labelled as a post-inspection sensitivity.

Phipson and Smyth explain why a randomly sampled permutation tail should not be reported as a zero p-value merely because no sampled draw is more extreme~\cite{phipson}. Hemerik and Goeman make the invariance and transformation assumptions of random-permutation testing explicit~\cite{hemerik}. These results motivate the plus-one tail convention and correct ordered relabelling in our simulation. They do not establish exchangeability of our token labels: that is an additional conditional modelling assumption. We accordingly report descriptive randomization comparisons rather than presenting them as an exact causal test of team skill.

\subsection{Uncertainty, period differences and the remaining research gap}
Clopper and Pearson provide exact binomial confidence limits~\cite{clopper}. Here those limits describe uncertainty in simulated exceedance counts, conditional on the archive, algorithm and sampling law. Repeating seeds checks Monte Carlo stability; it cannot quantify data-collection error, guarantee executable prices or remove hindsight selection.

K\"unsch develops resampling methods for dependent observations~\cite{kunsch}; Politis and Romano develop the stationary bootstrap for weakly dependent stationary series~\cite{politis}. These papers motivate preserving dependence when constructing uncertainty summaries. Our creation-hour block resampling is a simpler, explicitly specified conditional procedure, not an implementation of the stationary bootstrap. Short selected windows do not establish stationarity, independence between blocks or transportability of the intervals to another market period.

The resulting research gap is specific. Existing studies provide evidence about organized pumps, mechanisms of token fraud, related-address inference, AMM execution and multiple-search bias. Our study brings these concerns together to describe following opportunities across independently reconstructed Pump.fun windows, with explicit cash locking and comparable random controls. Table~\ref{tab:literature} summarizes the distinctions. We do not claim that related work contains no trading analysis, or that these selected windows identify a causal relationship between the SOL price and memecoin activity. The intended contribution is a reproducible comparison of conditional opportunity estimates and their failure modes across periods.

\begin{table}[htbp]\centering\small
\caption{Selected literature and its relation to the present research question. A source's fraud labels, sampling frame or trading result is not transferred to our dataset.}
\begin{tabular}{p{0.22\linewidth}p{0.31\linewidth}p{0.37\linewidth}}
\toprule
Research strand & What is established or studied & Consequence for this study\\\midrule
Organized pumps~\cite{kamps,xu,li_pump,dhawan,hamrick,lamorgia} & Event detection, information asymmetry and trading around coordinated announcements. & Separate a price pattern, prior knowledge and an outsider's available rule.\\[0.4em]
Wash trading and scam tokens~\cite{wash,xia,cong,cernera} & Transaction structures and mechanism-specific scam indicators. & Funding reachability alone does not verify those mechanisms.\\[0.4em]
Pump.fun and Solana~\cite{factories,chen_sol,li_sol,mongardini} & Manipulation mechanisms, behavioral identification and early outcome prediction. & Specify signal time, outcome definition, platform and execution assumptions separately.\\[0.4em]
Address inference~\cite{meiklejohn,victor} & Linking pseudonymous addresses under chain-specific assumptions. & Treat descendant sets as candidate teams with unmeasured identity error.\\[0.4em]
AMM execution~\cite{angeris,daian,lehar,qin} & Constant-product pricing and ordering-dependent trading opportunities. & Include finite size and costs; distinguish a quote or mark from a fill.\\[0.4em]
Backtest selection~\cite{sullivan,white,hansen,bailey,harvey} & Performance assessment after searching many alternatives. & Disclose the search, use matched controls and separate earlier selection from later outcomes.\\[0.4em]
Conditional inference~\cite{phipson,hemerik,clopper,kunsch,politis,holm} & Randomization, binomial limits and dependence-aware uncertainty. & State the randomization assumptions and the limited target of every interval.\\
\bottomrule\end{tabular}\label{tab:literature}
\end{table}
\FloatBarrier

\paragraph{Contribution.} The closest studies of Pump.fun and Solana scams classify manipulation or predict fraudulent tokens from early trading~\cite{factories,chen_sol,li_sol}; they do not follow identified creator groups through finite-capital accounts. This article is a logical continuation of our previous work \cite{gafarov2026}. We contribute three things. First, a period-by-period protocol that reports the same account in three information regimes: hindsight team thresholds, cash instead of marks and online qualification. Second, a reserve-state pipeline for bonding-curve accounts (own deposit in the counterfactual pool, units, closing states, strict boundary). Third, ten independently reconstructed windows, one of which was fixed before its results were computed, which show how the apparent opportunity differs across market periods and why. We do not claim to detect manipulation, to identify owners or to demonstrate a deployable strategy.

\section{Data, units and admissible paths}

All data had been obtained directly from the Solana blockchain by using getBlock API method and stored in PostgreSQL databases. Decoding Solana blockchain transactions is a challenging task, especially when nested transactions are involved. As a result, the decoded data on token price dynamics may contain inaccuracies. The measurement cohort contains non-Mayhem launches with a full archive hour remaining and an admissible state in their first minute. Full-hour archive coverage is not evidence that every transaction was collected. Eligibility is an ex-post observable-data condition, including in the midpoint exercise. A launch can have a full archive hour but a shorter admissible reserve path. Missing paths do not receive invented zero returns.

The dataset fields are treated as bonding-curve account balances in SOL. A configured rent-exempt balance is removed before any range, direction or continuity check. Solana's storage minimum depends on account size~\cite{solana_accounts}. The reference classes are 0.00123192, 0.00193488 and 0.00194184 SOL for the configured 49-, 150- and 151-byte layouts. The default class follows creation time and window; this is not a census of historical account configurations. At a documented initial buy, a positive pre-balance compatible with exactly one class can identify that token's offset without using its outcome or its tested post-reserve. A zero pre-balance before initial account allocation has economic pre-reserve zero only when the initial-buy flag and creation timestamp agree. The data-quality tables disclose the evidence available for these assignments.

Earlier archive balances are interpreted at their stored float32 centres. A quantisation interval containing protocol zero uses zero as its representative; an interval wholly below zero is invalid. An above-boundary value is never projected down to 85, and numerical uncertainty at that boundary is separately counted. With $R$ the normalized real reserve, every admitted pre- and post-state satisfies $0\le R\le85$. Buys cannot reduce it and sells cannot increase it; zero changes are allowed. Recorded \texttt{price\_sol}, token quantities and reported trade SOL do not choose primary states or value trades. Amount consistency is examined only in a declared sensitivity.

The 85-SOL boundary is a conservative analysis domain retained for comparability, not a claim that the protocol can never record a greater reserve. The documented reference initial reserves and real-token supply imply completion at approximately 85.005359 SOL under the ideal constant-product expression~\cite{pump}. The new source contains completion events just above 85. Such states remain outside our chosen domain and are censored without clipping; this alone does not classify them as corrupt. The same conservative limit also applies after adding the follower's own deposit. The domain restriction may therefore remove opportunities near genuine completion.

Within each chronological timestamp and recorded slot, pre-to-post states define a directed multigraph. A connected Euler trail with one start and one end has an identified closing state even if the internal order is not unique. A balanced multi-state cycle requires a known prior anchor. Unanchored cycles, inconsistent degrees and disconnected components cannot supply a primary quote for that block. This determines closing states without choosing an arbitrary signature order. It verifies internal consistency of supplied states, not the completeness of the chain history.

A block that cannot be resolved (an unanchored cycle, inconsistent degrees, disconnected components, a row with a negative or non-finite reserve, or a row whose reserve change contradicts its side) supplies no state and is skipped. A continuity break from the preceding closing state is repaired: the first internally consistent closing state after the break becomes the new anchor, and the trades in between are unknown. Only a reserve above 85 SOL, which leaves the model domain, ends a path at its recorded timestamp, and a path is not reopened afterwards. A completed-second state is available only at the following second boundary. Within-second peaks describe reserve reach and define seeds but cannot by themselves execute an exit. After a repair the monitor of the 85-SOL boundary sees only the recorded states, and a peak inside a gap is not seen. Skipped blocks and repaired gaps are counted by token.

All three quote variants share these units, closing states and validity boundaries. The primary retains every admitted completed second. The state-change variant retains seconds containing a reserve change of at least 0.01 SOL; the reported-SOL variant retains seconds whose constituent trade amounts all pass the declared consistency check. Retained seconds keep the same closing reserve, rather than a different last qualifying trade. The counterfactual-cap monitor uses the common primary history even when an entry or exit quote is filtered out.  Reserve paths are censored only where the state leaves the model domain (reserve above 85 SOL) or cannot be recovered. Gaps in the recorded trades are bridged and unresolved blocks are skipped, so that 0.6\% of the eligible paths of the nine earlier windows and 1\% of those of 2026-09 end before their hour. 

\section{Groups and the trading model}
\subsection{Funding groups and hindsight ranking}
We retain the source programme's family of directed SOL-funding descendants. Endpoint incidence is counted within each window and high-activity addresses are excluded. A seed has a positive maximum reserve during the first six seconds and subsequently an observed maximum greater than 60 SOL and at least twice that early maximum. Root creators and directed descendants define candidate wallet sets. Wallet Jaccard similarity strictly above 80 percent prunes near-duplicates before the activity filter, in creation-time, mint and identifier order. Candidate groups have between eleven and 399 known launches with completed hours; ranking separately requires more than ten eligible tokens. The upper activity bound and pruning order retain the source method's selection universe. A seed's criterion is not a token-price multiple. The exit grid is 4, 6, and successive even real-SOL levels through 84, followed by 85. The standardized ranking principal is the minimum of 0.01 SOL, one percent of entry reserve and remaining headroom below 85; infeasible orders contribute a neutral factor. All other base settings are disclosed below.

\begin{table}[htbp]\centering\small\setlength{\tabcolsep}{4pt}
\caption{Parameters by status. Protocol: taken from the public programme documentation. Technical: conventions needed to define states and accounts. Chosen: research choices of this study, varied in the sensitivity analyses.}\label{tab:v8_params}
\begin{tabular}{p{.17\linewidth}p{.43\linewidth}p{.33\linewidth}}\toprule
Status & Parameter & Setting\\\midrule
Protocol & Reference virtual reserves and token supply; completion level & 30 SOL, $1.073\times10^9$ tokens; about 85.005 SOL\\
Technical & Admitted real-reserve domain; reference offset of virtual reserve & $0\le R\le85$; 30 SOL\\
Technical & State availability; maximum quote age; observation horizon & next second boundary; 30 s; one hour\\
Technical & Minimum order; cash stop & 0.001 SOL; 0.1 SOL\\
Technical & Funding graph: endpoint incidence, Jaccard pruning, group size & $\le1000$; $>0.8$ pruned; 11--399 launches\\
Chosen & Entry delay & 4 s (sensitivity 1 to 60 s, Sections~\ref{sec:deep} and~\ref{sec:strategies})\\
Chosen & Fee per side; model slippage & 1\% (1.25\% and 2\% scenarios); 0 (5\% scenario)\\
Chosen & Position: cash fraction, cap, reserve fraction & 10\%; 0.1 SOL; 1\%\\
Chosen & Exit threshold grid; ranking gate & 4, 6, \ldots, 84, 85 real SOL; index $>10$\\
Chosen & Selected teams; random cohorts & at most three; 10,000 random cohorts per account control\\
Chosen & Selection-matched nulls & 50,000 draws (token unit), 10,000 draws (creator unit); plus-one tail $(b+1)/(B+1)$\\
\bottomrule\end{tabular}\end{table}
Of these settings, the cost and the delay move the results most. With 2\% fee and 5\% slippage per side the main account of the same teams and thresholds ends between 0.77 and 8.93 SOL, above one SOL in 7 of 8 windows, with free cash between 0.09 and 6.79 SOL; the window-level numbers are saved with the extension tables of the release. The entry-delay sensitivity (1 to 60 seconds) and the cost scenarios do not change which windows have a team, since team selection uses the common base configuration, but they change the accounts.

For each token and reserve exit level $t$, $\rho_{it}$ is the finite liquidation-or-mark value divided by entry debit at the standardised notional. The group index is
\[
 I_g=\max_t\exp\left\{\sum_{i\in g}\log[1+w(\rho_{it}-1)]\right\}.
\]
The grid searches a common entry time and the disclosed reserve levels independently in every candidate group. A token without a feasible minimum order contributes a neutral factor. The index is a retrospective ranking statistic; it is not the wealth of a trader who can serially reinvest all marks. At most the three top-ranked qualifying groups form the leader union, with a shared token bought once. Selecting those identities uses the entire window in the hindsight experiments.

Here $w=0.1$ is the fraction of current wealth staked on each token, the same fraction as the cash fraction of the account. The product form is the growth of a trader who bets $w$ of wealth on each token in sequence and reinvests everything; its logarithm is additive, so a group's score is the sum of the logarithmic factors of its tokens, maximised over the exit threshold $t$. The requirement $I_g>10$ is the retained gate of the source method: a group must multiply the notional of such a trader by more than ten. Because the index ignores cash locking and lets a trader hold any number of simultaneous positions, it cannot be read as the wealth of an investor with limited capital. Table~\ref{tab:v8_index} gives an example with three tokens.

\begin{table}[htbp]\centering\small\setlength{\tabcolsep}{2.5pt}
\caption{Worked example of the ranking index with $w=0.1$ for the first three tokens of the 2024-04 team 1 at its best threshold of 54 real SOL. $\rho$ is the liquidation value of a standardised position divided by its entry debit; the factor $1+w(\rho-1)$ is the growth of a trader who stakes the fraction $w$ of current wealth on each token in turn; the last column is the running product. With all 13 tokens of the team the index is 60.47, above the threshold of 10.}\label{tab:v8_index}
\begin{tabular}{rrrrr}\toprule
Token & $\rho$ & Factor & Log factor & Running product \\\midrule
1 & 1.64 & 1.064 & 0.062 & 1.064 \\
2 & 1.81 & 1.081 & 0.078 & 1.150 \\
3 & 1.64 & 1.064 & 0.062 & 1.224 \\
\bottomrule\end{tabular}\end{table}

\subsection{Own deposit, finite proceeds and the strict boundary}
Use the declared constant-product reference $x_0=30$ virtual SOL, $y_0=1{,}073{,}000{,}000$ tokens and $k=x_0y_0$~\cite{pump}. For entry principal $a$ at historical real reserve $R_e$, let $X_e=30+R_e$. With $\sigma$ the scenario slippage wedge, normalized inventory is
\[
 u=\frac{q}{k}=\frac{a}{X_e(X_e+a)(1+\sigma)}.
\]
The historical reserve excludes the follower's deposit. At an admissible counterfactual state $R_x+a\le85$, net sale value is
\[
 g(u,R_x,a)=\min\left\{R_x+a,\frac{u(30+R_x+a)^2}{1+u(30+R_x+a)}\right\}(1-f)(1-\sigma).
\]
Entry debit is $a(1+f)$. An unchanged reserve returns $a$ before costs. Both the historical reserve and the counterfactual reserve must remain admissible; the virtual maximum is consequently 115 SOL. If the observed reference path first implies $R+a>85$, the position's executable path ends and its value freezes at the preceding admissible quote. Later recovery does not reopen it or return cash. No sale at an unobserved completion boundary is invented. The monitor covers the entire observed holding path, not just the planned exit state.
A worked example of the boundary on one real token is in Appendix~\ref{sec:app_boundary}.

The principal is limited by free cash, the configured cash fraction and position cap, the entry-reserve fraction, fees and headroom below 85. Cash locking is chronological across tokens; closes precede opens at equal times, followed by token identifiers. Minimum orders and the cash stop are enforced. The same constraints apply to leaders, random traders, sensitivity accounts and examples.

These expressions are conditional on a constant-product reference and the supplied reserve path with the added deposit. Other traders' responses, priority, transaction failure, migration and historical configuration changes are not replayed. The own-deposit correction does not make the archived market a fully reconstructed counterfactual market. Base fees and higher-cost/slippage cases are declared scenarios; they are not a period-by-period measurement of actual execution costs~\cite{pump,fees}.

\subsection{Exits, cash and marks}
The main exit is specific to each team: the position is closed at the first later completed-second closing state at or above the reserve threshold that maximised the team's own ranking index. This threshold is chosen on the whole window, so it is a hindsight choice, and it is kept unchanged when seed tokens are removed. A common 40-SOL exit and time-based exits are reported as alternatives. Entry can already exceed a threshold, so a completed trade is not necessarily profitable. The time-based policies combine a holding-time exit with an optional stop at half the entry reserve. A fresh uncensored historical zero reserve can close a position if its counterfactual pool still contains the follower's deposit. Stale, unavailable, historically censored or counterfactually censored exits cannot supply cash.

Unresolved positions are marked at the last admissible state before their applicable horizon. They remain locked and cannot finance another entry. We report cash, marked inventory and total equity separately. Assigning all unresolved inventory zero value is a stress endpoint, not a confidence bound or a measured terminal liquidation. For interpretation, the historical marginal-price ratio is $[(30+R_b)/(30+R_a)]^2$, not $R_b/R_a$; it is neither a recorded price multiple nor a finite portfolio return.

A numerical example of equity, cash and marks is in Appendix~\ref{sec:app_equity}.

\paragraph{Why marks are not counted as cash.} A mark is the model value of selling the position at its last admissible state, net of the fee, so within the model it is the proceeds of such a sale. The main account nevertheless never sells a position of this kind, because its only exit is the reserve threshold of the team; cash is therefore realised only when the threshold is reached. Whether a mark could be turned into cash is a separate question about execution: the pool may have lost its liquidity (for example after a sale by the creator), the recorded state may no longer be current when an order arrives, and the sale itself moves the price of a thin pool. For this reason marks are reported separately and are not added to the cash that finances entries. Appendix~\ref{sec:strategies} shows that selling every position that is still open at the end of the first hour at its last admissible state raises free cash and allows later entries; that result rests on the same assumption that such a sale executes at the recorded reserve, and it is reported as a sensitivity, not as the main account.

\section{Comparisons and uncertainty}
\subsection{Random traders, policies and seed removal}
The tokens-by-timing design crosses leader versus uniform random tokens with fixed versus random entry/exit timing, using the same opportunity count and cash rules within each comparison. Random-time entries are uniform in the first minute and attempted exits uniform afterwards through the first hour; unavailable attempts are rejected. A separate 100-opportunity control gives every window the same opportunity budget. Windows without leaders retain random controls but do not receive an artificial leader return.

Policy comparisons report the full leader union and the same identities after excluding seed-criterion tokens from leaders and the random universe. A further after-seed subset contains launches after the full seed criterion becomes observable, including the early-reserve doubling requirement. Because group identities and their funding memberships remain retrospectively selected, this subset is not a causal discovery test. Entry-delay sensitivity varies the launch delay; it does not measure actual network latency.

For the team-threshold experiment, each disjoint traded team cohort chooses the threshold maximizing its own ranking criterion on that very cohort, including separately after seed removal. Shared tokens follow the higher-ranked group. Random cohorts receive the same threshold grid and team chunk sizes, with a uniform ordered injection before splitting. This is an in-sample threshold search conditional on chosen cohorts, not a mathematical upper bound on cash-ledger wealth or a reproduction of the prior graph discovery search.

\subsection{Selection-matched nulls}
The token null uses one shared injection within creation-hour by seed-status strata for all candidate groups, preserving sizes and overlaps while repeating the candidate/threshold search and qualification gate. The creator null injects entire creators with an exactly equal joint creation-hour by seed-token histogram. Thus group token counts, temporal/seed composition, creator overlaps and within-creator sequences are preserved. Singleton strata may leave creators immovable; coverage is disclosed rather than concealed by coarsening sizes. Both nulls condition on the discovered graph and membership structure and do not rerun funding-graph discovery.

The conservative plus-one tail includes ties: with $B$ draws and $b$ draws at least as extreme, $p=(b+1)/(B+1)$. The token null uses $B=50{,}000$ and the creator null $B=10{,}000$, so the smallest attainable values are about $2\times10^{-5}$ and $10^{-4}$; the 10,000 of the parameter table is the number of random cohorts per account control. Mid-$p$ is diagnostic only: it is not automatically a conservative p-value~\cite{midp,hemerik,phipson}. A fixed family contains both index statistics, both exchangeable units and all ten usable windows. We report Holm~\cite{holm} across this full family and the disclosed within-statistic window families; tied or sparse windows are not deleted after inspecting their tail fractions. These are conditional randomization comparisons under the stated exchangeability assumptions, not calibrated guarantees for the full research process. Repetitions are split across deterministic seeds. Exact binomial intervals describe Monte Carlo tail counts~\cite{clopper}, while creation-hour block bootstraps describe conditional per-token means~\cite{kunsch,politis}; neither establishes input truth or fillability.

\subsection{Temporal separation and parameter sensitivity}
At each window's midpoint the graph, exclusions and ranking use only earlier information; a training token must have completed its hour. Group identities and the portfolio policy are frozen before evaluating later known wallets. Separately, a post-inspection sensitivity expands funding knowledge causally: the knowledge time of a wallet is the minimum possible maximum edge time along a directed funding path. A link must be observable strictly before a later creator's launch. These exercises remain conditional on the observable-data cohort and are not cross-window training tests.

Funding caps and transfer thresholds, reserve-retention filters, capital fractions, position caps, fees and slippage are varied independently within each period. The full surfaces are supplied. No best financial parameter or hindsight threshold is promoted to an independently validated strategy.

\section{Results by independent window}
\subsection{Selected teams and the main account}\label{sec:accounts}
\begin{figure}[htbp]\centering\includegraphics[width=\linewidth]{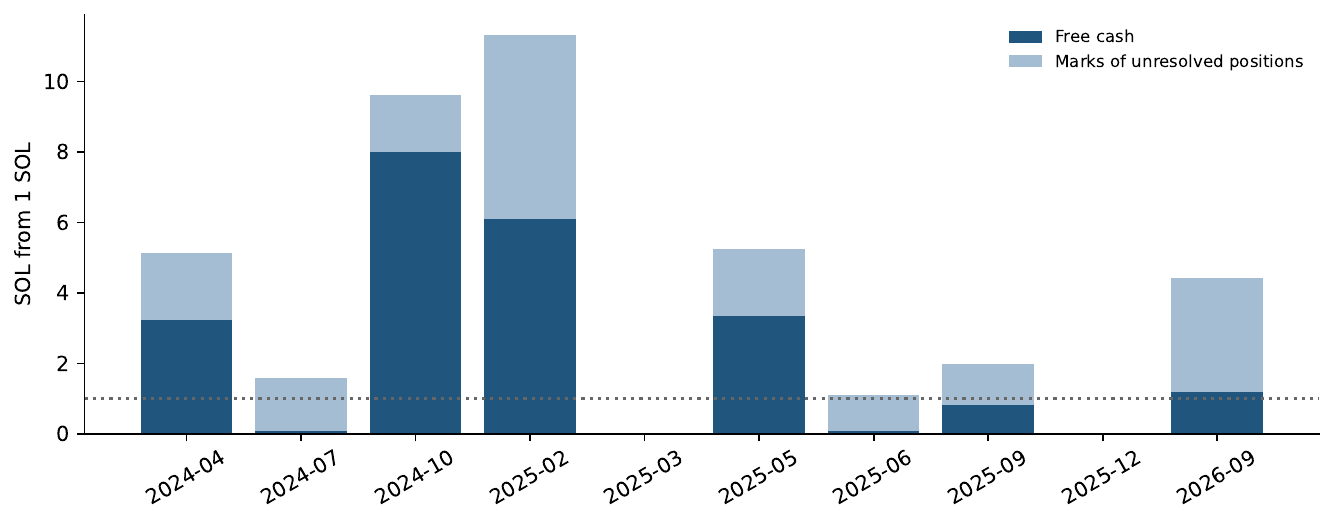}\caption{Cash and marks of the main account in each window. Bars split final equity into free cash and the marks of unresolved positions; the dotted line is the initial capital. Windows without a qualifying team are blank.}\label{fig:v8_accounts}\end{figure}

In the main account every selected team exits at its own best reserve threshold, a hindsight choice that maximises the team's ranking index on its own tokens. Shared tokens follow the higher-ranked team, and one account per window starts with one SOL (see. Table \ref{tab:v8_accounts} and Figure \ref{fig:v8_accounts}). Equity ends above one SOL in all 8 windows with a team, between 1.10 and 11.32 SOL, while the random cohorts that run the same threshold search have medians of at most 1.06 SOL. Free cash exceeds one SOL in 5 windows (2024-04, 2024-10, 2025-02, 2025-05, 2026-09). In the windows with less than one SOL of free cash (2024-07, 2025-06, 2025-09) most of the value is in positions that never reached their threshold: marks are 58--94\% of equity there and 17--94\% across all windows (Table~\ref{tab:v8_decomp}). The ledgers of closed trades give the realised part; marks are model values and cannot finance new entries. 

\begin{table}[htbp]\centering\scriptsize\setlength{\tabcolsep}{2.5pt}
\caption{Main account of each window: every selected team exits at its own best reserve threshold (chosen in hindsight on the full window), shared tokens follow the higher-ranked team, one account starts with 1 SOL. Equity is cash plus marks of positions that never reached their threshold; bought/sold counts positions opened and closed. Random: median equity of random cohorts of the same size that run the same threshold search, and the number of 10,000 random accounts at least as large. Dashes: no qualifying team.}\label{tab:v8_accounts}
\begin{tabular}{lrrrrrrrrr}\toprule
Window & Teams & Tokens & Equity & Cash & Marks & Bought/sold & Open & Random med. & Random $\ge$ \\\midrule
2024-04 & 2 & 142 & 5.14 & 3.22 & 1.92 & 137 / 19 & 118 & 0.99 & 0/10,000 \\
2024-07 & 3 & 379 & 1.59 & 0.10 & 1.49 & 64 / 3 & 61 & 0.92 & 70/10,000 \\
2024-10 & 3 & 64 & 9.61 & 8.00 & 1.61 & 64 / 25 & 39 & 1.06 & 0/10,000 \\
2025-02 & 3 & 276 & 11.32 & 6.09 & 5.23 & 265 / 95 & 170 & 0.92 & 0/10,000 \\
2025-03 & 0 & -- & -- & -- & -- & -- & -- & -- & -- \\
2025-05 & 3 & 84 & 5.26 & 3.35 & 1.91 & 84 / 17 & 67 & 0.94 & 0/10,000 \\
2025-06 & 3 & 389 & 1.10 & 0.10 & 1.01 & 110 / 54 & 56 & 0.87 & 194/10,000 \\
2025-09 & 1 & 83 & 1.99 & 0.83 & 1.16 & 83 / 17 & 66 & 0.94 & 7/10,000 \\
2025-12 & 0 & -- & -- & -- & -- & -- & -- & -- & -- \\
2026-09 & 1 & 189 & 4.43 & 1.20 & 3.24 & 140 / 21 & 119 & 0.89 & 0/10,000 \\
\bottomrule\end{tabular}\end{table}

\paragraph{Why only part of the union is bought.} A team's tokens are not all traded (Table~\ref{tab:v8_funnel}). In 2 windows the largest part of the union is skipped because free cash had already fallen below the 0.1 SOL stop: the account holds many positions that wait for a threshold that most tokens never reach, and every purchase locks cash. This is a property of an exit that waits for a high reserve, not of data availability. Tokens without an admissible entry state are rare.

\FloatBarrier

\paragraph{Seed tokens.} The main account trades every token of a selected team, seed tokens included. Seed status is defined by the outcome of a token, so a test that removes all seed tokens (Appendix~\ref{sec:seed}) is a diagnostic of how concentrated the result is, not a trading scenario; it understates what a follower could earn and does not undo the hindsight choice of teams and thresholds.

\subsection{From a discovered team to a possible signal}\label{sec:prospective}
\begin{figure}[htbp]\centering\includegraphics[width=\linewidth]{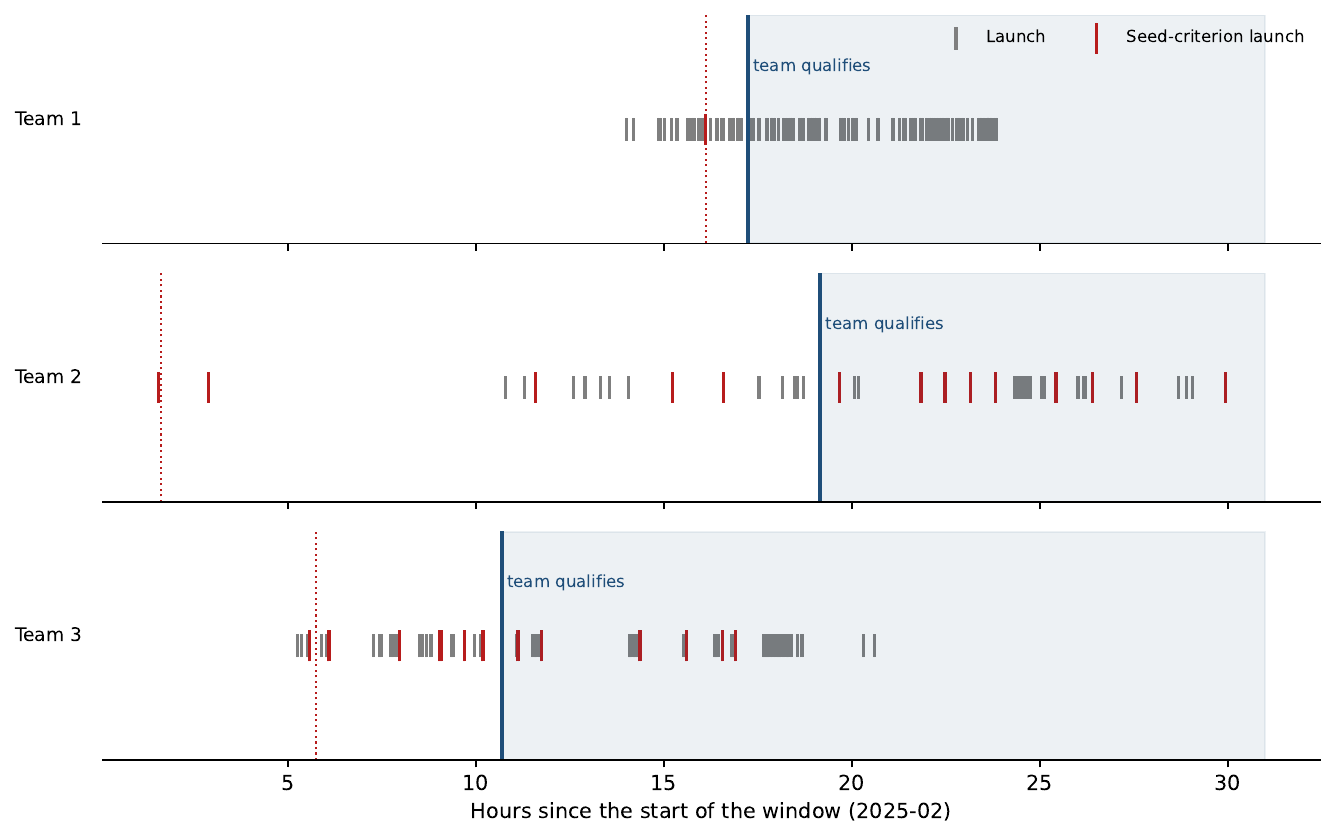}\caption{Timeline of the three teams of the 2025-02 window. Ticks are the team's launches (red: seed criterion met). The dotted line marks the first observable seed event, the solid line the moment when the online ranking index first exceeds ten; the shaded region contains the launches that could have been followed.}\label{fig:v8_timeline}\end{figure}

A follower needs a signal before a launch, not a ranking at the end of the window. Figure~\ref{fig:v8_timeline} shows the sequence for the three teams of 2025-02: launches of each team on the time axis, the first observable seed event, and the moment at which the online index first exceeds ten using tokens whose hour had completed. Only launches after that moment could have been followed. Table~\ref{tab:v8_prospective} applies the rule to every team. 9 of the 19 teams qualify only when their last launch completes, so no later launch is available: their high indices in the hindsight account reflect launches that could not have been followed under this rule. The other 10 teams leave later launches; their separate accounts end above one SOL in 10 cases with free cash above one SOL in 7. 9 of the 10 accounts exceed at least 95\% of random later launches with the same threshold. The threshold used is the one that was best on the completed tokens, not the full-window threshold. Funding links are taken from the full window, so the table is optimistic about when links were observable: future edges and future wallet membership may influence the composition of a team, and the table must not be read as a signal that could have been traded in real time.

The per-team table (Table~\ref{tab:v8_prospective}) and the comparison of frozen and expanding identity sets are in Appendix~\ref{sec:app_prospective}.
\FloatBarrier

\subsection{What a team is in the data}\label{sec:structure}
\begin{figure}[htbp]\centering\includegraphics[width=\linewidth]{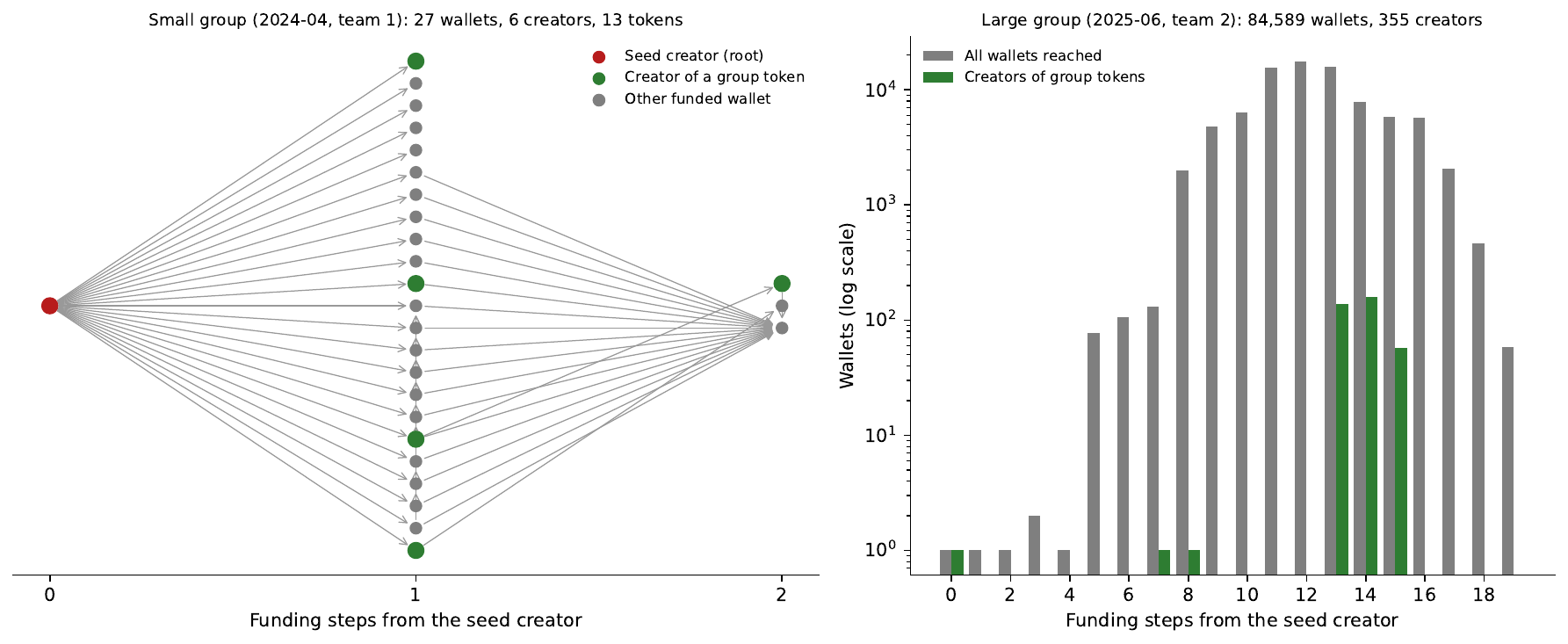}\caption{Funding structure of two real groups (addresses anonymised). Left: the seed creator, its funded wallets and the creators of the group's tokens. Right: wallets reached and token creators by number of funding steps in the large group.}\label{fig:v8_structure}\end{figure}

A team is a set of wallets reachable from a seed creator by native SOL transfers, after excluding high-activity endpoints. Figure~\ref{fig:v8_structure} shows two real groups with addresses anonymised. The small group of 2024-04 has a clear fan-out: the seed creator funds 32 wallets directly, and 67\% of the 6 creators of its tokens are funded by one wallet. The large group of 2025-06 is a different object: 84,589 wallets are reachable in up to 19 transfer steps, its 355 token creators sit 7--15 steps from the seed creator, and no single wallet funds more than 0.3\% of them. Such a group is a long dispersal tree whose leaves launch tokens, and it is much more consistent with common infrastructure (a distributor or service) than with a coordinated team. The ranking and the account treat both objects in the same way; we therefore describe teams as candidate groups and do not draw inferences about ownership.

\begin{figure}[htbp]\centering\includegraphics[width=.92\linewidth,height=.78\textheight,keepaspectratio]{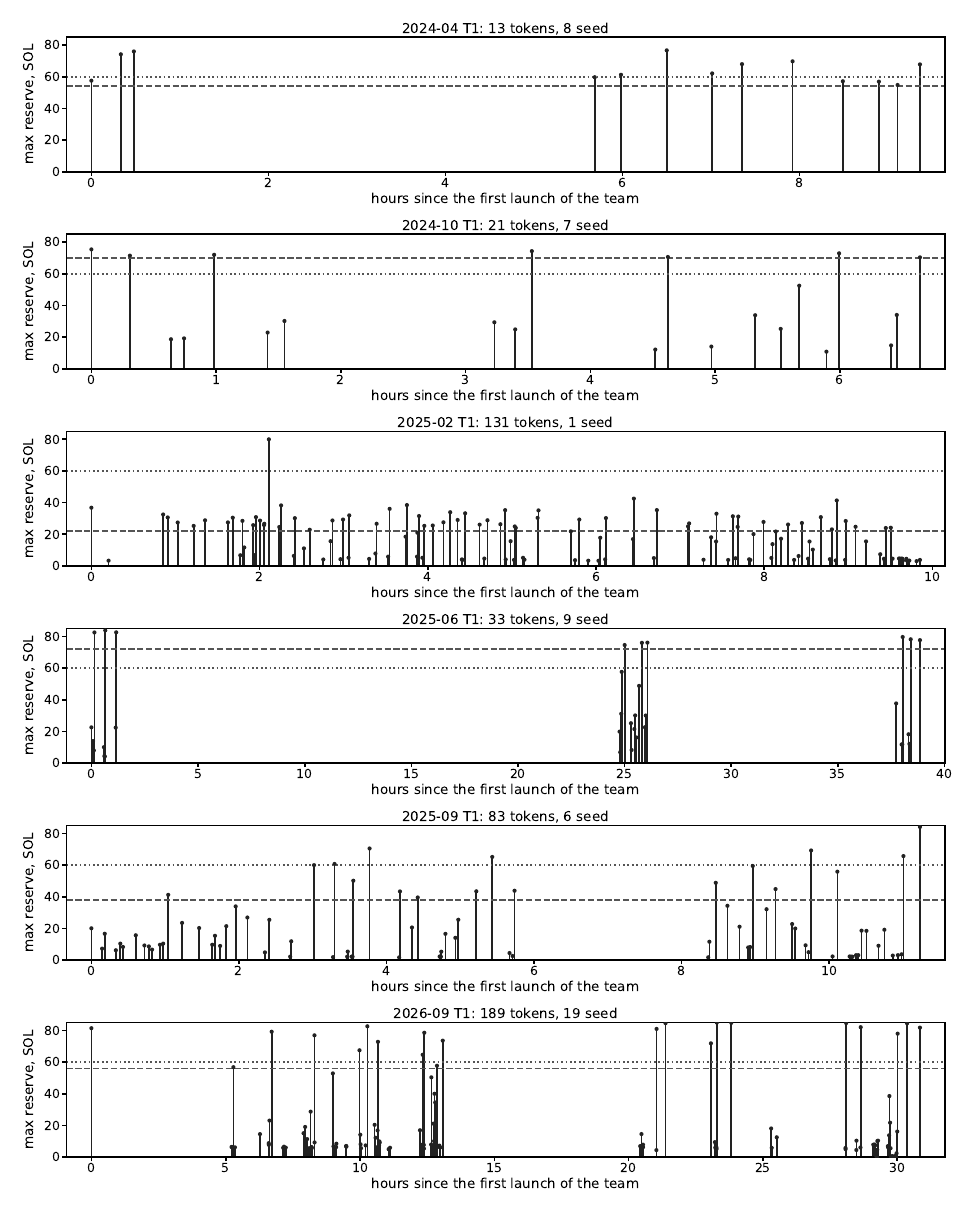}\caption{Launch dynamics of the top-ranked team (T1) in six of the windows. Each stem is one token: its maximum valid reserve in SOL against the hours since the team's first launch (each panel has its own time axis; the vertical axis is cut at 85 SOL). The dotted line is 60 SOL, the dashed line the team's best threshold; the panel title gives the number of tokens and of seed tokens.}\label{fig:v8_dynamics}\end{figure}

Figure~\ref{fig:v8_dynamics} shows the launches of the top-ranked team in six of the windows with the maximum reserve of every token on the time axis. Three features stand out. First, the pattern differs strongly between teams: the two earliest windows shown (2024-04, 2024-10) are short sequences in which many launches are seed tokens, while the top team of 2025-02 launches 131 tokens of which only one reaches the seed level. Second, launches come in bursts separated by pauses of many hours (2025-06), and the seed tokens often fall inside one burst; the groups are therefore not steady producers. Third, many seed tokens end only a few SOL above the 60 SOL line, close to the team's threshold, so the outcome of a token depends on a narrow margin. The panels show the ranked teams after the fact: they illustrate what the data contain and are not a forecast of what a team will launch next.

\FloatBarrier

\paragraph{Worked example.} One team is followed trade by trade, with every trade listed, in Appendix~\ref{sec:example}.
\FloatBarrier
\subsection{A team followed trade by trade}\label{sec:example}
\begin{figure}[htbp]\centering\includegraphics[width=\linewidth]{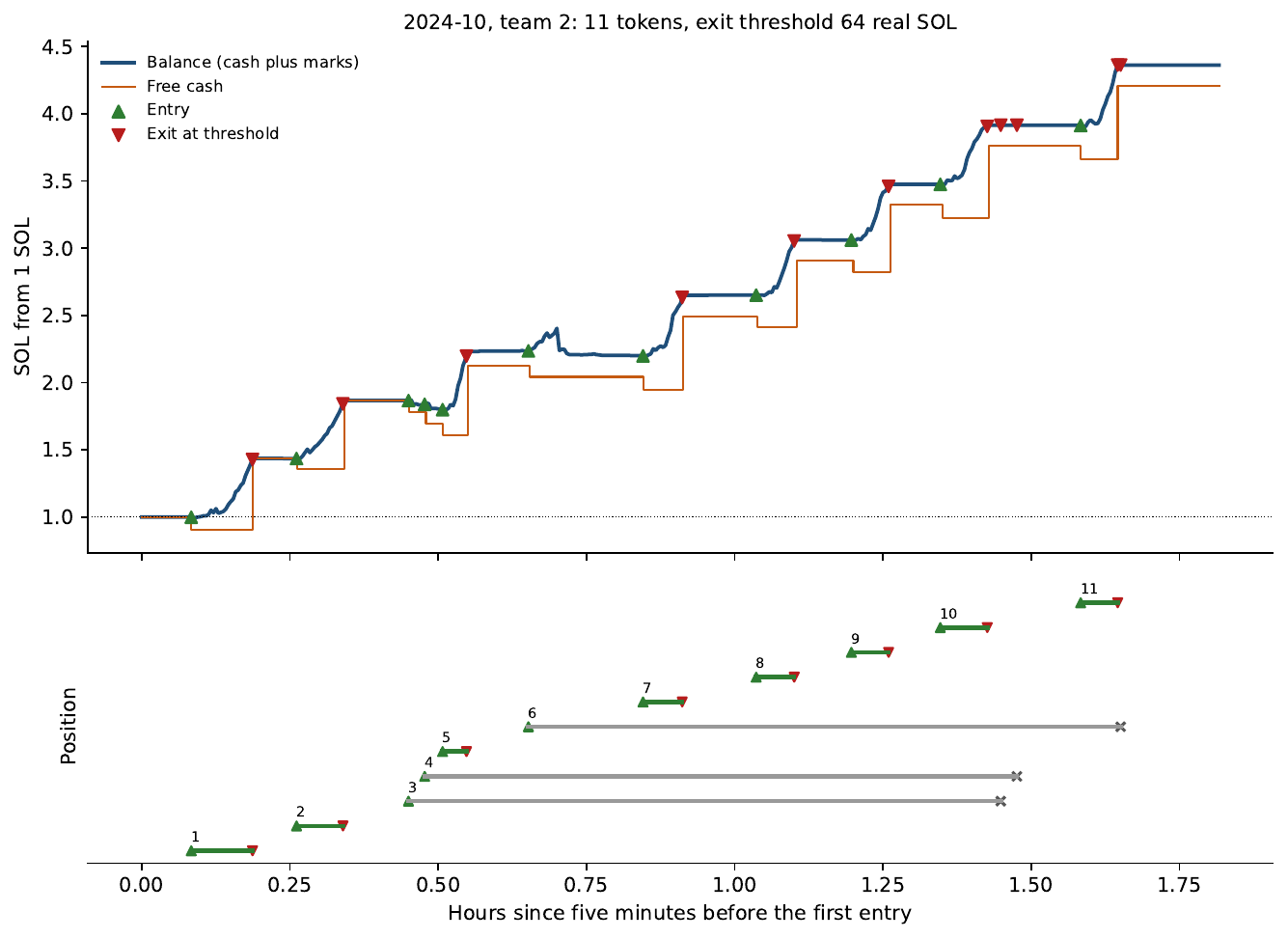}\caption{Balance, free cash, entries and exits of the example team (2024-10, team 2). Entries and exits are marked on the balance curve; the lower panel shows each position from entry to exit or to the end of its hour.}\label{fig:v8_example}\end{figure}

Figure~\ref{fig:v8_example} follows one team through its trades: the balance as cash plus marks, free cash, and each entry and exit. The team is the second team of 2024-10 (11 tokens, threshold 64 real SOL); it is a favourable example chosen to show the mechanics, not a typical window. 8 of the 11 positions reach the threshold after a few minutes and release cash; 3 stay open and end near the empty-pool mark. Table~\ref{tab:v8_example} lists every trade with the reserves at entry and exit, the size and the SOL released.

\subsection{Sensitivity of the discovery of a team}
Whether a team is found at all is more sensitive than the account of a team that is found. In 2026-09 the highest index among candidates is 596.6 with native SOL transfers and 5.9 when owner-attributed WSOL edges are added, which is below the gate of 10, so the number of selected teams changes from 1 to 0. Adding edges changes both reachability and endpoint incidence, and therefore the membership of every candidate; we have not isolated which links cause the change. The endpoint-incidence cap and the minimum transfer size change the number of qualifying groups in the same window between 1 and 5. Reserve-quote filters (state-change and reported-SOL variants) rebuild the admissible states and can likewise qualify a group that the primary quotes do not; these cases are listed in the appendix. The existence of a team is therefore a result of the whole specification, not a stable property of the window.
\FloatBarrier
\subsection{Windows compared}\label{sec:compare}
Table~\ref{tab:v8_windows} places the ten windows side by side. Sorting them by what happens to the main account gives the following patterns: realised (5); marks only (3); no qualifying team (2). The comparison below does not assign a cause.

\emph{Cash versus marks.} Equity exceeds one SOL in every window with a team, but the share held in marks ranges from 17\% to 94\%. The windows with cash above one SOL (2024-04, 2024-10, 2025-02, 2025-05, 2026-09) are those in which the net result of the closed trades exceeds the cost of the positions still open; in the others the account is full of positions waiting for a threshold that most tokens never reach (Tables~\ref{tab:v8_decomp} and~\ref{tab:v8_funnel}).

The windows also differ in length (24 to 79 hours), in launches and in the number of teams. The table therefore supports a comparison of patterns, not a ranking of market conditions.

\begin{table}[htbp]\centering\scriptsize\setlength{\tabcolsep}{2.5pt}
\caption{The ten windows side by side: recorded length (hours), launches, share of eligible tokens whose reserve path is censored before their hour ends, the archive's own price-regime label (no price is measured in this study), teams, equity and cash of the main account, later tokens in the strict frozen / causally expanding midpoint tests, and the pattern. Pattern: realised means free cash above 1 SOL.}\label{tab:v8_windows}
\begin{tabular}{lrrrlrrrrl}\toprule
Window & Hours & Launches & Censored  & Teams & Equity & Cash & Later (strict / exp.) & Pattern \\\midrule
2024-04 & 24 & 11,916 & 0.5\% &  2 & 5.14 & 3.22 & 0 / 0 & Realised \\
2024-07 & 31 & 14,282 & 1.2\% &  3 & 1.59 & 0.10 & 13 / 20 & Marks only \\
2024-10 & 33 & 22,398 & 0.8\% & 3 & 9.61 & 8.00 & 0 / 0 & Realised \\
2025-02 & 31 & 73,394 & 1.0\% &  3 & 11.32 & 6.09 & 0 / 525 & Realised \\
2025-03 & 24 & 25,856 & 0.5\% &  0 & -- & -- & 0 / 173 & No qualifying team \\
2025-05 & 55 & 57,939 & 0.4\% &  3 & 5.26 & 3.35 & 0 / 252 & Realised \\
2025-06 & 79 & 84,821 & 0.5\% &  3 & 1.10 & 0.10 & 4 / 293 & Marks only \\
2025-09 & 71 & 81,233 & 0.6\% &  1 & 1.99 & 0.83 & 0 / 0 & Marks only \\
2025-12 & 72 & 52,377 & 0.4\% &  0 & -- & -- & 0 / 0 & No qualifying team \\
2026-09 & 48 & 83,742 & 1.1\% &  1 & 4.43 & 1.20 & 0 / 0 & Realised \\
\bottomrule\end{tabular}\end{table}
\FloatBarrier

\section{Two windows in detail and parameter sensitivity}\label{sec:deep}
We examine one favourable and one unfavourable window in detail and vary every analysis parameter one at a time. The windows are chosen by their own results, so they are extreme cases, not a representative sample.

\subsection{Choice of the windows}
Eight of the ten windows have a qualifying team. The best window is 2024-10: it has the highest free cash (8.00 SOL, 83\% of its equity). The window with the highest equity, 2025-02 (11.32 SOL), is not chosen because its free cash is lower (6.09 SOL; 46\% of its equity is marks). The worst window is 2025-06: it has the lowest equity (1.10 SOL), cash near the stop (0.10 SOL) and 91\% of its equity in marks. Windows without a team (2025-03, 2025-12) have nothing to vary.

\subsection{Anatomy of the two accounts}
\begin{figure}[htbp]\centering\includegraphics[width=.62\linewidth]{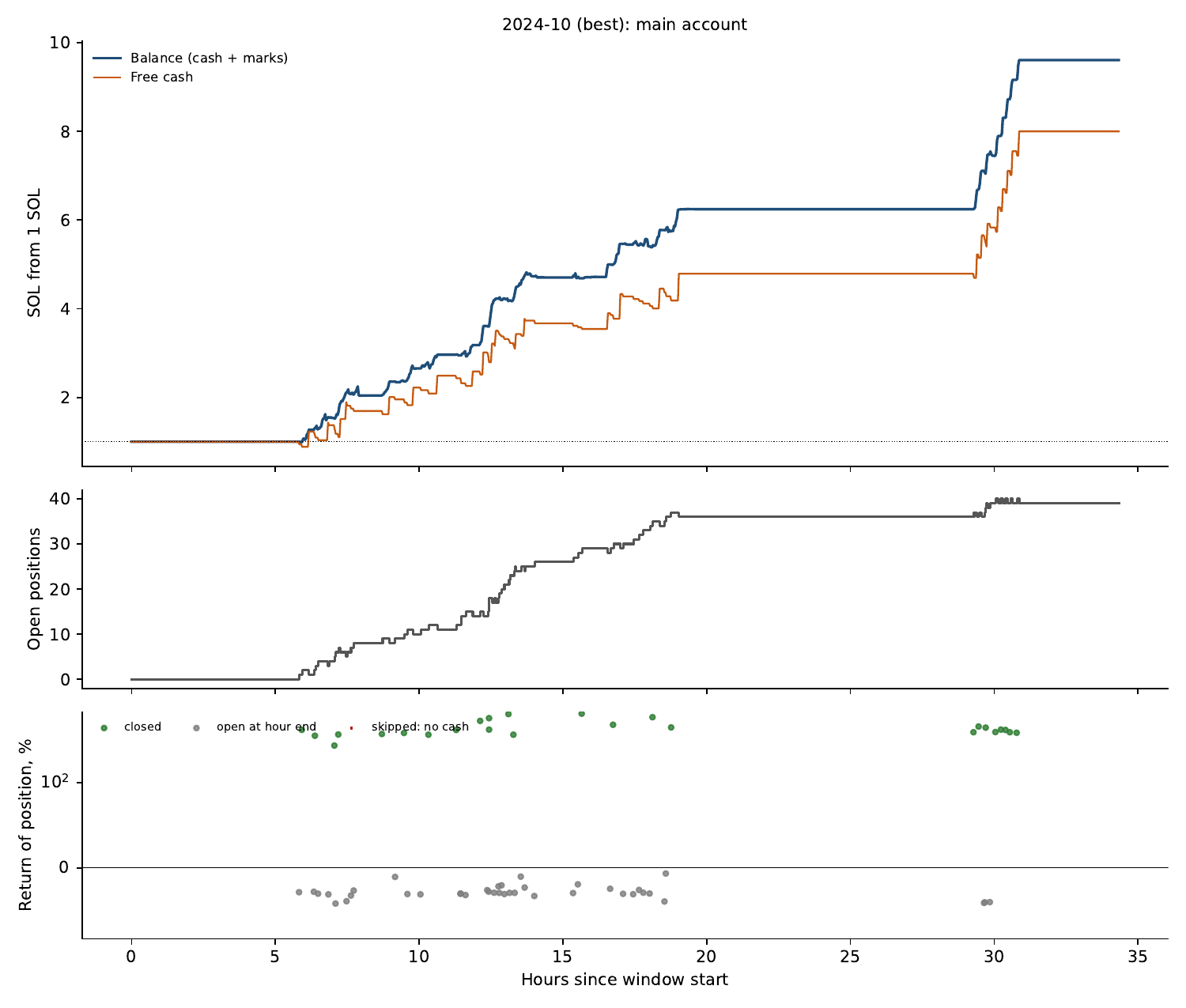}\hfill\includegraphics[width=.62\linewidth]{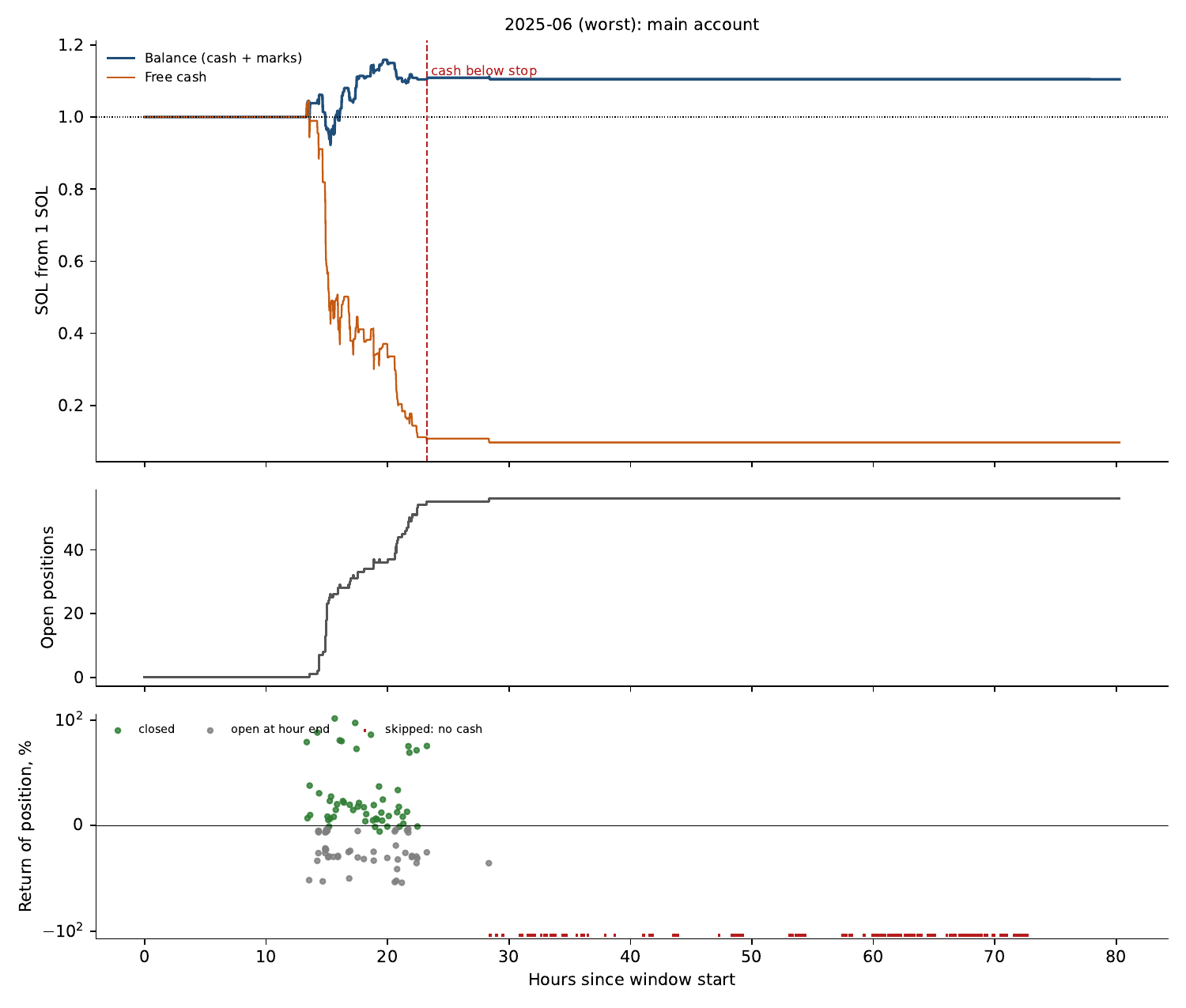}\caption{Balance, free cash, open positions and the return of every position in the best (top) and the worst (bottom) window. The dashed line marks the moment when free cash falls below the 0.1 SOL stop; red ticks are tokens of the union that were not bought for lack of cash.}\label{fig:deep_anatomy}\end{figure}

Table~\ref{tab:deep_compare} and Figure~\ref{fig:deep_anatomy} compare the accounts position by position. In 2024-10 the three teams are compact (11 to 32 tokens, one token per creator, thresholds of 56 to 70 SOL). The 25 positions that reach the threshold close after a median of 13.5 minutes with returns between +408\% and +761\% (5th to 95th percentile); the 39 positions that stay open end at a median of -29\% and sum to -0.68 SOL. The 24 seed positions produce +8.98 of the +8.61 SOL total; the other 40 positions together produce -0.37 SOL. In 2025-06 the three teams have 33, 356, 19 tokens and thresholds of 72, 16, 72 SOL; the largest has 355 creators and 84,589 wallets, and one team owns no token of its own because its tokens belong to higher-ranked teams. The account buys 110 tokens between hour 13.3 and hour 23.2 of the window, then free cash falls below the stop, 56 positions stay open for the rest of the window and 279 further tokens of the union are not bought. The 54 closed positions return a median of +18\% after 0.2 minutes, while 100\% of the open positions lose value.

\begin{table}[htbp]\centering\small\setlength{\tabcolsep}{2.5pt}
\caption{The best and the worst window in comparison (main account, team thresholds). Equity, cash and marks in SOL from one SOL; closed and open positions at the end of the first hour; returns are relative to the entry debit.}\label{tab:deep_compare}
\begin{tabular}{lrr}\toprule
Quantity & 2024-10 (best) & 2025-06 (worst) \\\midrule
Equity / cash / marks & 9.61 / 8.00 / 1.61 & 1.10 / 0.10 / 1.01 \\
Equity without seed tokens & 0.94 & 1.05 \\
Union tokens / positions opened & 64 / 64 & 389 / 110 \\
Closed / open at hour end & 25 / 39 & 54 / 56 \\
Median hold of closed positions, min & 13.5 & 0.2 \\
Median return, closed / open & +482\% / -29\% & +18\% / -25\% \\
Share of seed tokens in the total result & 104\% & 54\% \\
Tokens skipped, cash below stop & 0 & 279 \\
Team thresholds, SOL & 70, 64, 56 & 72, 16, 72 \\
Tokens per team & 21, 11, 32 & 33, 356, 19 \\
Censored share of team paths & 0\%, 0\%, 0\% & 3\%, 0\%, 0\% \\
\bottomrule\end{tabular}\end{table}

\FloatBarrier

\paragraph{Parameter sweeps.} Every analysis parameter was varied alone in both windows and in the six other windows with a team; the sweeps and their pooled results are in Appendix~\ref{sec:deep_sweeps}.

\paragraph{Limits.} Every value was chosen after inspecting the data, the team and its threshold are defined on the whole window, and the random control repeats the threshold search but not the search for groups. The two sweeps that change the result most in 2025-06, the stop and the graph variant, were chosen after the result was known.

\section{Variations of the main rule}\label{sec:strategies_summary}
Appendix~\ref{sec:strategies} changes one rule of the main account at a time in the eight windows with a team. Entering after one second instead of four raises the geometric mean equity from 3.82 to 4.20 SOL, but in only three windows; the gain is not systematic. Position size has no consistent best value, and without the cap of one percent of the entry reserve the results become bets on a few tokens. 73\% of the positions that the main accounts open stay open at the end of their hour and hold 67\% of the invested SOL; selling them at the last admissible mark raises the geometric mean equity from 3.82 to 5.65 SOL, because it turns marks into cash. All thresholds are chosen in hindsight, and no rule was tested in the prospective setting of Section~\ref{sec:prospective}.

\section{Interpretation and limitations}
The question that motivates the study is whether the right teams could have been followed. The answer is layered. With teams and thresholds chosen on the whole window, every window with a team shows an apparent gain, and random controls that run the same search rarely match it (between 0 and 194 of 10,000 random accounts reach the team account). That gain is largely marks and, when the ranking is recomputed online, disappears for the teams that qualify only after their last launch. For the teams that can be followed, the accounts that remain are smaller and are mostly marks. These findings describe conditional retrospective opportunities: the funding links, the ranking, the thresholds and the observation cohort all use the full window, and the model adds one follower's deposit to a reference curve without reconstructing rival responses, priority, failed inclusion or migration. The detailed analysis of the best and the worst window (Section~\ref{sec:deep}) adds a caution: in the worst window the account stops trading after hour 23 because its cash is locked in open positions, so a change of the exit rule or of the funding graph alone can raise the result several-fold. Parameters that are best when the 8 windows with a team are pooled are minimal costs and delay; others, such as position size, have no consistent best value. A prospective test would fix the discovery rules, the thresholds and the data schema before collecting a new sample (Section~\ref{sec:future}).\par

\paragraph{Three different statements.} The results support three statements of different strength and they should not be merged. (i) A large model balance: with teams and thresholds chosen on the whole window the main account ends above one SOL in every window with a team, but this is a hindsight figure and mostly marks. (ii) A result above random portfolios: random cohorts that run the same threshold search rarely match the team accounts, which shows that the selected tokens differ from random tokens under the same exit search. (iii) Statistical confirmation: only 3 of the 40 tests of the selection-matched nulls are rejected after the Holm adjustment. These nulls keep the discovered funding graph and permute token or creator labels; they do not rerun the group search, the filtering, the threshold optimisation and the capital accounting, so they test a conditional hypothesis and do not calibrate the whole selection procedure. Nulls that repeat the construction and selection of the graph on permuted data would be a stricter test and are left to future work; we state this as a limitation, not as a result.
The strongest defensible interpretation is conditional: selected historical groups can produce attractive model accounts in some separately observed windows, but selection, data quality, incomplete cash realization and reference-mechanism assumptions remain material. A cleaned subset cannot identify returns on all launches. Large in-sample balances and small conditional tail fractions do not establish the availability of those identities before trading.

The windows were chosen for interest, differ in duration. Observation quality may therefore explain part of a period difference. We do not annualize these balances, infer population trends or identify a causal SOL-price effect. Native-SOL comparisons do not equal dollar returns or equalize liquidity and trading opportunities. A future prospective study would predeclare windows, verify historical units/configurations and fees, discover groups causally and reserve genuinely future observations. This study supplies separate-window descriptions and auditable assumptions for that work.

\section{Directions for further research}\label{sec:future}
The results of this study are conditional retrospective accounts. Six directions would turn them into evidence that can be tested, and each has a hypothesis that can fail.

\paragraph{1. Causally correct identification.} At every moment the funding graph, the membership of wallets, the ranking and the signals should be rebuilt only from data observed up to that moment, and the result compared with the retrospective network used here. \emph{Hypothesis:} the share of the hindsight result that survives is the information that a follower could actually have used; we expect it to be smaller than the online figures of this paper, because funding links are taken from the full window there. \emph{Test:} replay each window in event time, with the graph updated edge by edge, and report the lag between the first observable link and the moment the team is recognised.

\paragraph{2. Prospective validation.} The discovery rules, the parameters, the data schema, the evaluation intervals and the full list of metrics should be fixed before a new sample is collected, and the policy should then be tested without change. \emph{Hypothesis:} if the teams are persistent producers, the team accounts will exceed random cohorts on the new sample under the frozen rules; if the apparent opportunity comes from selection, they will not. \emph{Test:} a registered protocol with a pre-specified sample size and one primary metric, the free-cash result of the online account, with the remaining metrics reported as secondary.

\paragraph{3. Validation of the economic entities.} A funding group is a graph construct and not an owner. The groups should be checked against external labels of services and exchanges, against a labelled subset of wallets, against the depth and the gaps of the funding edges, and by transfer type (native SOL, wrapped SOL, seed-based transfers). \emph{Hypothesis:} compact fan-out groups behave like coordinated teams, whereas long dispersal trees correspond to shared infrastructure, as the two groups of Section~\ref{sec:structure} suggest. \emph{Test:} repeat the account separately for the two kinds of groups and for the groups that remain after the removal of labelled service wallets.

\paragraph{4. Real execution.} The model adds one follower's deposit to a reference curve. A realistic test should include the fees of the historical configuration, failed and delayed inclusion, competition from other buyers, migration of liquidity, and the risk of positions that cannot be sold. \emph{Hypothesis:} the free-cash results of the accounts with many open positions are the most sensitive to execution, because their value is mainly marks that rely on a sale at the recorded reserve. \emph{Test:} small live or simulated-with-latency accounts compared with the model account for the same tokens.

\paragraph{5. Statistical evaluation of the full search.} The selection-matched nulls of this study keep the discovered graph. Nulls that repeat the graph construction, the filtering, the threshold optimisation and the capital accounting on permuted data would separate the conditional uncertainty that we report from the full uncertainty of the procedure. \emph{Hypothesis:} the full-procedure null will reject in fewer windows than the conditional null. \emph{Test:} rerun the entire pipeline on each permuted sample and compare the distributions of the final statistics.

\paragraph{6. Market regimes.} The windows should be extended and the results related to liquidity, activity, the structure of the market and the SOL price, without causal conclusions that the design does not support. \emph{Hypothesis:} the concentration of the result in seed tokens and the share of marks differ systematically between periods of high and low creation activity. \emph{Test:} a longer panel of windows with a pre-specified regime variable, and a comparison that keeps the pipeline unchanged.

\section*{Data and code availability}
The data and replication code that support the findings of this study are available from the  author upon reasonable request.

\clearpage\appendix
\section{A stress test without seed tokens}\label{sec:seed}
A seed token is one whose observed reserve passes 60 real SOL and at least doubles its first-six-second maximum. Teams are born from seeds: a creator with a seed token starts a candidate group, and the ranking index is then evaluated on all tokens of the group, including the seeds. The main account trades every token of a selected team, seeds included; this is possible because group membership is defined by funding links, not by the outcome of a token. As a stress test, Table~\ref{tab:v8_seed} keeps the teams and thresholds and drops the seed tokens. Equity then exceeds one SOL in 3 of 8 windows (2025-05, 2025-06, 2025-09), and free cash stays at or below 0.37 SOL in every window.

\paragraph{How this test should be read.} It is a diagnostic of how concentrated the result is, not a trading scenario. Seed status is defined by the outcome of a token (its reserve later passes 60 SOL), so removing all seeds also removes tokens that a follower who had recognised the team could have traded; the test therefore understates what a follower could earn and must not be read as the return of a strategy. It also does not undo the retrospective choice of teams and thresholds, and the removed tokens are exactly those on which the groups were found. Hence the test shows that the gain is concentrated in few tokens, but not that the teams have no value. Section~\ref{sec:prospective} instead recomputes the ranking online. Refitting the threshold on the remaining tokens  gives equity of 1.06 to 4.25 SOL, above one SOL in 8 of 8 windows, but this too is in-sample and mostly marks.

\begin{table}[htbp]\centering\scriptsize\setlength{\tabcolsep}{2.5pt}
\caption{Contribution of the seed tokens. Same teams and thresholds with and without the tokens that meet the seed criterion (an observed maximum reserve above 60 real SOL and at least twice the first-six-second maximum); the threshold is not re-optimised. Columns: tokens, equity and cash without and with seeds, net result of the closed seed trades, closed net result of the full and seed-free accounts, and random accounts at least as large as the seed-free account.}\label{tab:v8_seed}
\begin{tabular}{lrrrrrrrrrr}\toprule
Window & Tok. all & Tok. no seed & Eq. all & Eq. no seed & Cash all & Cash no seed & Seed net & Net all & Net no seed & Random $\ge$ \\\midrule
2024-04 & 142 & 127 & 5.14 & 0.95 & 3.22 & 0.10 & 3.28 & 4.27 & 0.01 & 0/10,000 \\
2024-07 & 379 & 352 & 1.59 & 0.88 & 0.10 & 0.09 & 0.73 & 0.73 & 0.00 & 123/10,000 \\
2024-10 & 64 & 40 & 9.61 & 0.94 & 8.00 & 0.09 & 8.98 & 9.29 & 0.30 & 0/10,000 \\
2025-02 & 276 & 248 & 11.32 & 0.77 & 6.09 & 0.09 & 7.51 & 12.26 & 0.00 & 0/10,000 \\
2025-05 & 84 & 69 & 5.26 & 1.22 & 3.35 & 0.10 & 4.27 & 4.87 & 0.60 & 0/10,000 \\
2025-06 & 389 & 363 & 1.10 & 1.05 & 0.10 & 0.10 & 0.06 & 0.57 & 0.48 & 10/10,000 \\
2025-09 & 83 & 77 & 1.99 & 1.53 & 0.83 & 0.37 & 0.46 & 1.21 & 0.75 & 0/10,000 \\
2026-09 & 189 & 170 & 4.43 & 0.88 & 1.20 & 0.10 & 4.61 & 4.97 & 0.28 & 192/10,000 \\
\bottomrule\end{tabular}\end{table}
\FloatBarrier

\section{Online qualification: per-team results and identity sets}\label{sec:app_prospective}
\begin{table}[htbp]\centering\scriptsize\setlength{\tabcolsep}{2.5pt}
\caption{Prospective qualification. The ranking index is recomputed online from tokens whose first hour had completed; a team qualifies at the first completion that leaves more than ten eligible tokens and an index above ten (hours from the start of the window). The follower then trades the team's later launches with the threshold that was best on the completed tokens (in parentheses: the full-window threshold of the hindsight account). Funding links are taken from the full window. Each team has its own 1-SOL account; random: median of random later launches with the same threshold and the number at least as large. No later launches means the team was recognisable only when its last launch had completed.}\label{tab:v8_prospective}
\begin{tabular}{lrrrrrrrrrrr}\toprule
Window & Team & Tokens & Qualifies (h) & Done & Later & Threshold & Equity & Cash & No seed & Random & $\ge$ \\\midrule
2024-04 & 1 & 13 & 19.4 & 11 & 0 & -- & -- & -- & -- & -- & -- \\
2024-04 & 2 & 129 & 20.1 & 107 & 20 & 80 (78) & 1.29 & 0.90 & 0.95 & 0.94 & 90/3,000 \\
2024-07 & 1 & 11 & 26.5 & 11 & 0 & -- & -- & -- & -- & -- & -- \\
2024-07 & 2 & 356 & 21.1 & 209 & 134 & 34 (84) & 1.03 & 0.09 & 0.86 & 0.88 & 371/3,000 \\
2024-07 & 3 & 12 & 7.2 & 11 & 0 & -- & -- & -- & -- & -- & -- \\
2024-10 & 1 & 21 & 17.7 & 12 & 6 & 70 (70) & 1.83 & 1.62 & 0.93 & 0.97 & 3/3,000 \\
2024-10 & 2 & 11 & 31.8 & 11 & 0 & -- & -- & -- & -- & -- & -- \\
2024-10 & 3 & 32 & 12.3 & 19 & 10 & 56 (56) & 1.46 & 1.15 & 0.87 & 0.95 & 27/3,000 \\
2025-02 & 1 & 131 & 17.2 & 20 & 99 & 24 (22) & 2.44 & 1.09 & 2.44 & 0.90 & 0/3,000 \\
2025-02 & 2 & 48 & 19.2 & 15 & 30 & 52 (50) & 3.62 & 2.77 & 1.29 & 0.87 & 0/3,000 \\
2025-02 & 3 & 97 & 10.7 & 34 & 59 & 78 (78) & 2.64 & 1.06 & 0.78 & 0.81 & 0/3,000 \\
2025-05 & 1 & 32 & 49.8 & 26 & 0 & -- & -- & -- & -- & -- & -- \\
2025-05 & 2 & 18 & 29.0 & 18 & 0 & -- & -- & -- & -- & -- & -- \\
2025-05 & 3 & 34 & 50.7 & 34 & 0 & -- & -- & -- & -- & -- & -- \\
2025-06 & 1 & 33 & 67.4 & 29 & 0 & -- & -- & -- & -- & -- & -- \\
2025-06 & 2 & 356 & 65.6 & 276 & 71 & 14 (16) & 1.33 & 0.55 & 1.32 & 0.84 & 0/3,000 \\
2025-06 & 3 & 19 & 68.2 & 19 & 0 & -- & -- & -- & -- & -- & -- \\
2025-09 & 1 & 83 & 53.3 & 51 & 32 & 38 (38) & 1.52 & 1.05 & 1.24 & 0.92 & 1/3,000 \\
2026-09 & 1 & 189 & 17.0 & 71 & 118 & 52 (56) & 3.76 & 2.26 & 0.73 & 0.78 & 0/3,000 \\
\bottomrule\end{tabular}\end{table}

\paragraph{Frozen and expanding identity sets.}  The frozen design keeps only the wallets that belonged to a team at the midpoint, so a later token qualifies only if its creator was already in that set. The expanding design adds a creator when a funding path from the team becomes observable strictly before that creator's launch. In 2025-02 the frozen set contains 0 later tokens while the expanding set contains 525: the difference is consistent with creators who are funded by a team wallet only after the midpoint, links that did not exist when the frozen set was fixed; we have not traced each of these creators. The expanding design is a post-inspection sensitivity; the strict design has later-token coverage in only 2 of the ten windows.
\FloatBarrier

\section{Parameter sweeps in the two windows and pooled over windows}\label{sec:deep_sweeps}
Each parameter is varied alone, with the others at their baseline: entry delay (1 to 60 s), quote age, fee (0 to 3\%), slippage (0 to 10\%), cash fraction, position cap, reserve fraction, cash stop, shift of every team threshold (plus or minus six grid steps), a common threshold for all tokens (all 42 grid values), a holding limit combined with a stop on the entry reserve ($5\times4$ grid), time-only exits, the number of teams, the index gate, nine funding-graph variants and three quote variants. Every setting is evaluated for all tokens and, as a stress test only (Section~
ef{sec:seed}), without the seed tokens, with a random control of the same size that receives the same threshold search (3,000 draws). The baseline account of each window equals the release account to $10^{-9}$.

\begin{figure}[htbp]\centering\includegraphics[width=\linewidth]{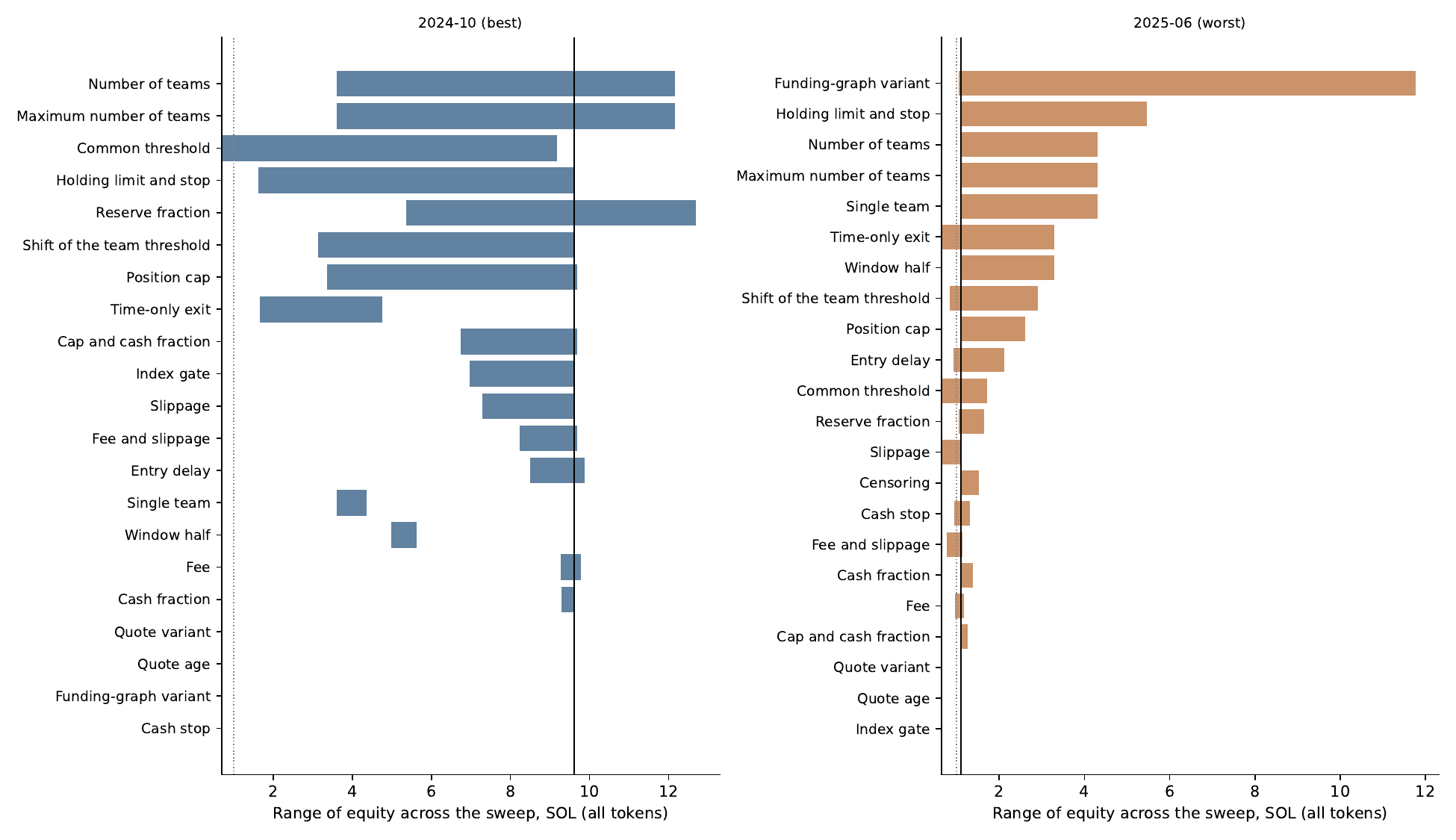}\caption{Range of equity (all tokens) over the sweep of each parameter in the best (left) and the worst (right) window. The vertical line is the baseline account, the dotted line one SOL.}\label{fig:deep_tornado}\end{figure}

The largest ranges of equity are, in 2024-10, maximum number of teams (3.60 to 12.16), number of teams (3.60 to 12.16), common threshold (0.70 to 9.17), holding limit and stop (1.62 to 9.61), and in 2025-06, funding-graph variant (1.06 to 11.76), holding limit and stop (1.10 to 5.47), maximum number of teams (1.10 to 4.31), number of teams (1.10 to 4.31) (Figure~\ref{fig:deep_tornado}). In 2024-10 the result is insensitive to the fee and to a delay of up to 30 seconds (equity stays between 9.27 and 9.88 SOL), because a few positions with a multiple of their entry reserve decide it; it is sensitive to the team threshold (a shift of one step upwards lowers equity from 9.61 to 7.69) and, as a scale, to the position size (cap 0.02 SOL: 3.37; reserve fraction 5\%: 12.69). In 2025-06 almost all ordinary parameters leave equity between 0.92 and 2.62 SOL. Two choices raise the result far above the baseline: a stop on the entry reserve (baseline 1.10, stop at 0.75 of the entry reserve 5.16), which frees the cash locked in the 100\% of losing open positions, and the funding-graph variant (activity cap 500: 11.76 with 708 tokens, because a different set of groups is selected). In 2024-10 all nine graph variants select the same three teams. Both choices were made after seeing the data; the median over the 20 holding-stop combinations in 2025-06 is 4.76 SOL.

\begin{figure}[htbp]\centering\includegraphics[width=.8\linewidth]{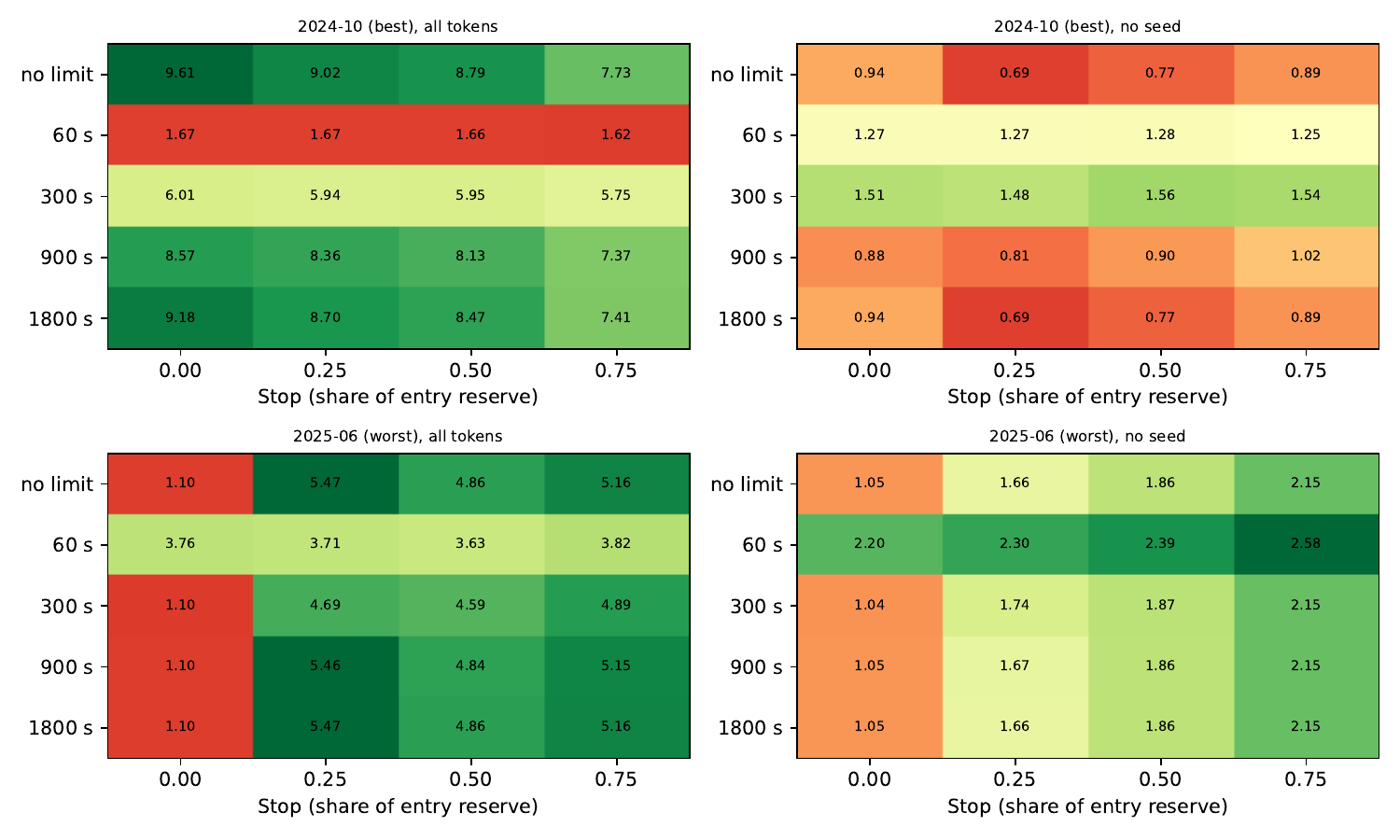}\caption{Equity for the team threshold combined with a holding limit and a stop on the entry reserve, all tokens and without seed tokens, in both windows.}\label{fig:deep_holdstop}\end{figure}

\paragraph{Threshold chosen on the first half of a team's tokens.} A threshold chosen on the first half of each team's tokens and applied to the second half gives 2.24 SOL for team 1 in 2024-10, 3.10 SOL for team 2 in 2024-10, 1.79 SOL for team 3 in 2024-10, against 2.24, 3.10, 1.98 with the full-window threshold; in 2025-06 2.41 (team 1), 2.27 (team 2), 2.26 (team 3) against 2.41, 2.25, 2.26. In these 6 team cases the early threshold differs from the full-window threshold by between -0.20 and +0.02 SOL; with so few tokens this is not a test of the threshold rule.

\subsection{Pooled over the 8 windows}
\begin{figure}[htbp]\centering\includegraphics[width=\linewidth]{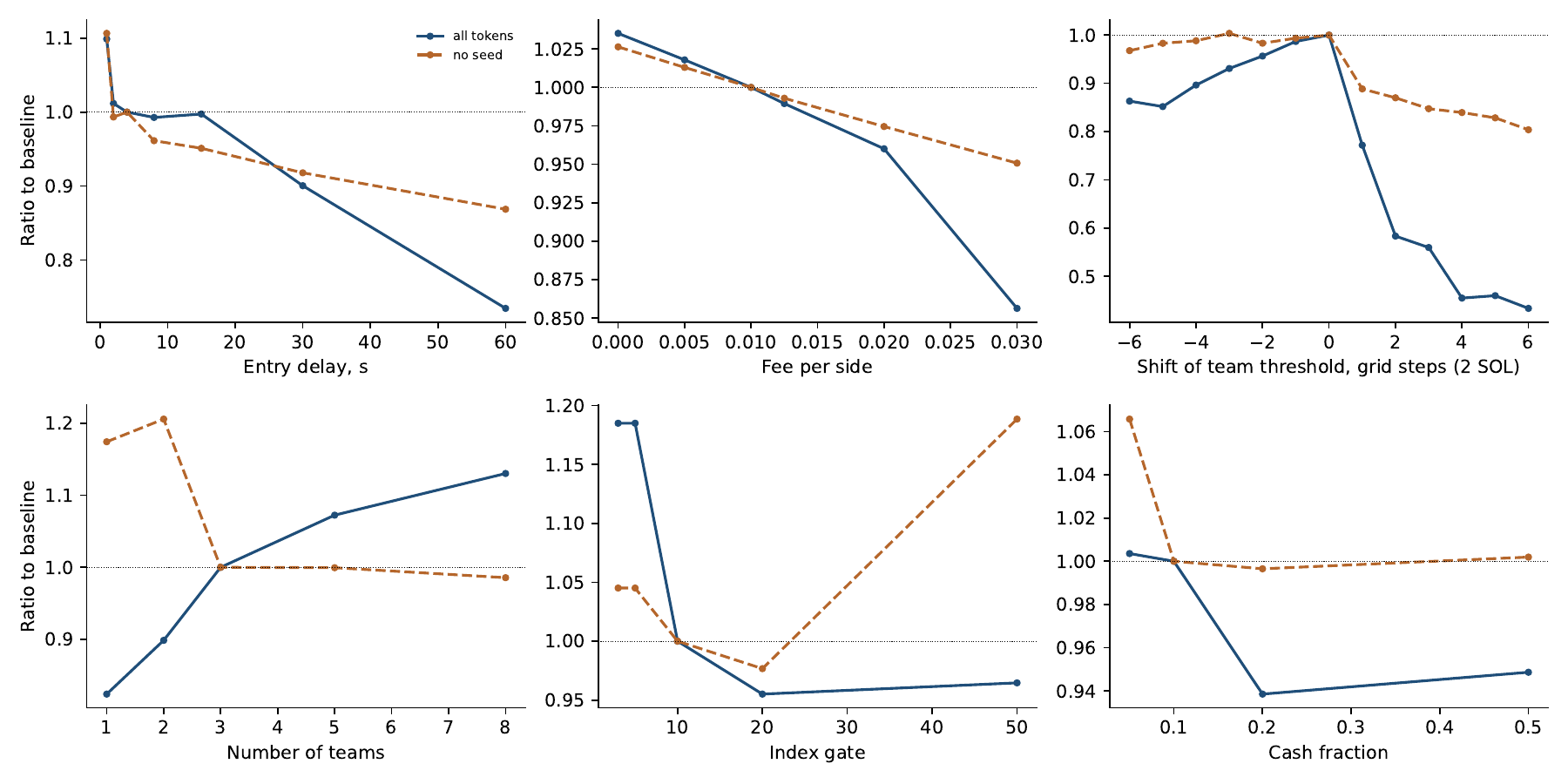}\caption{Geometric mean over windows of the ratio of equity to the baseline setting, for all tokens (solid) and without seed tokens (dashed).}\label{fig:deep_cross}\end{figure}

The same sweeps were run in the 6 other windows with teams. Table~\ref{tab:deep_pooled} and Figure~\ref{fig:deep_cross} show, for each parameter, the setting with the highest geometric mean ratio of equity to the baseline. Lower fees are better in all 8 windows (zero fee: ratio 1.04), and zero slippage is best everywhere. An entry delay of one second beats four seconds in 3 of 8 windows for all tokens and in 7 without seeds. Without seed tokens the pattern is toward faster and smaller exits: a holding limit with a stop (best setting 60/0.0, ratio 1.30), a single team instead of three (ratio 1.21) and thresholds shifted downwards. For all tokens the team threshold itself is best, and more teams raise the pooled ratio (ratio 1.13 for 8 teams) but in only 2 of 8 windows. Position size, cash stop and quote age have no consistent best value. These ratios compare equity, not returns, so in windows with equity below one SOL a ratio above one means smaller losses.

\begin{table}[htbp]\centering\footnotesize\setlength{\tabcolsep}{2.5pt}
\caption{Parameter values that are best when the sweeps of 8 windows are pooled. For each parameter and token set: the setting with the highest geometric mean ratio of equity to the baseline setting, that ratio, and the number of windows in which the setting beats the baseline. Baseline: fee 1\%, delay 4 s, team thresholds, three teams, index gate 10, 10\% cash fraction, cap 0.1 SOL, reserve fraction 1\%, stop 0.1 SOL. All values were chosen after inspecting the data.}\label{tab:deep_pooled}
\begin{tabular}{lrrrrrr}\toprule
Parameter & All: best & Ratio & Beats & No seed: best & Ratio & Beats \\\midrule
Fee per side & 0.0 & 1.04 & 8/8 & 0.0 & 1.03 & 8/8 \\
Entry delay, s & 1.0 & 1.10 & 3/8 & 1.0 & 1.11 & 7/8 \\
Holding limit (s) / stop & 0/0.25 & 1.20 & 3/8 & 60/0.0 & 1.30 & 4/8 \\
Shift of team threshold, steps & -1 & 0.99 & 1/8 & -3 & 1.00 & 3/8 \\
Common threshold, SOL & 56.0 & 0.78 & 0/8 & 4.0 & 1.22 & 7/8 \\
Time exit only, s & 60 & 0.61 & 3/8 & 60 & 1.20 & 4/8 \\
Maximum teams & 8 & 1.13 & 2/8 & 2 & 1.21 & 1/8 \\
Index gate & 3.0 & 1.18 & 3/8 & 50.0 & 1.19 & 2/5 \\
Graph variant & ex2000\_sol0.2 & 1.36 & 3/8 & ex2000\_sol0.2 & 1.48 & 3/8 \\
Cash fraction & 0.05 & 1.00 & 3/8 & 0.05 & 1.07 & 3/8 \\
Position cap, SOL & 1.0 & 1.01 & 3/8 & 0.02 & 1.10 & 4/8 \\
Reserve fraction & 0.02 & 1.03 & 4/8 & 0.02 & 1.01 & 3/8 \\
Cash stop, SOL & 0.05 & 1.05 & 2/8 & 0.2 & 1.01 & 5/8 \\
\bottomrule\end{tabular}\end{table}

\FloatBarrier

\section{Strategy variations: entry speed, position size, locked capital and exit at a collapse}\label{sec:strategies}
This section changes one rule of the main account at a time in the eight windows with a team (team thresholds, entry after 4 s, fee 1\% per side, stop at 0.1 SOL of cash, the 85-SOL boundary, own deposit in the pool). The simulator was extended with fixed-amount and capital-fraction sizing, an exit when the reserve falls from its peak, and a sale of the positions still open at the end of the first hour. In the base case such positions are never sold, only marked. The extended simulator reproduces the engine of the main account to $10^{-12}$ when the rules are switched off and an independent Python replica on 35 accounts of the study to $2\cdot10^{-15}$; these checks cover arithmetic, not the archived reserves. For the exit rules the threshold of every team is chosen again on the whole window, as in the main method, and the random controls repeat that search; every number below is therefore an in-sample value selected in hindsight. Without seed tokens the baseline depends on whether the thresholds are kept (1.00, as in the main account, the entry rows and the sizing rows) or refitted on the seed-free tokens (1.62, as in the other rules); the refitted value is again a hindsight choice. Table~\ref{tab:strat_rules} summarises the rules.

\begin{table}[htbp]\centering\footnotesize\setlength{\tabcolsep}{3pt}
\caption{Rules for entry, release of locked capital and exit at a collapse, pooled over the 8 windows with a team. Geometric mean (over windows) of equity in SOL from 1 SOL, number of windows with equity above one SOL, the smallest window value, and the number of windows in which the rule is at least as good as the v8 baseline. Team thresholds are re-chosen for every rule on the whole window (hindsight), so they are in-sample values. Entry rows keep the baseline thresholds, so their baseline without seed tokens is the row for 4 s (0.98), not the refitted value of the first row (1.58).}\label{tab:strat_rules}
\begin{tabular}{p{4.4cm}rrrrrrrr}\toprule
Rule & All: geo.\ mean & $>1$ & Worst & Beats v8 & No seed: geo.\ mean & $>1$ & Worst & Beats v8 \\\midrule
v8 baseline (no release, no collapse rule) & 3.82 & 8/8 & 1.10 & 8/8 & 1.62 & 8/8 & 1.06 & 8/8 \\
Entry after 4 s (thresholds kept) & 3.82 & 8/8 & 1.10 & 8/8 & 1.00 & 3/8 & 0.77 & 8/8 \\
Entry after 1 s & 4.20 & 8/8 & 1.97 & 3/8 & 1.11 & 5/8 & 0.78 & 7/8 \\
Entry after 2 s & 3.87 & 7/8 & 0.99 & 3/8 & 1.00 & 2/8 & 0.77 & 6/8 \\
Sell all open positions at the hour end & 5.65 & 8/8 & 1.99 & 8/8 & 1.86 & 7/8 & 0.89 & 7/8 \\
Collapse $-50\%$ of reserve, peak $\ge0$ & 3.94 & 8/8 & 1.33 & 3/8 & 1.60 & 7/8 & 0.59 & 4/8 \\
Collapse $-50\%$, peak $\ge8$ SOL & 4.39 & 8/8 & 1.54 & 3/8 & 1.59 & 8/8 & 1.06 & 5/8 \\
Collapse $-50\%$, peak $\ge8$ SOL, 2 s latency & 4.27 & 8/8 & 1.54 & 3/8 & 1.57 & 8/8 & 1.06 & 4/8 \\
Collapse $-50\%$, peak $\ge8$ + sell at hour end & 4.88 & 8/8 & 1.54 & 3/8 & 1.83 & 7/8 & 0.90 & 5/8 \\
Stop at 50\% of the entry reserve & 4.18 & 8/8 & 1.96 & 3/8 & 1.61 & 7/8 & 0.58 & 3/8 \\
\bottomrule\end{tabular}\end{table}

\subsection{Faster entries}
\begin{figure}[htbp]\centering\includegraphics[width=.9\linewidth]{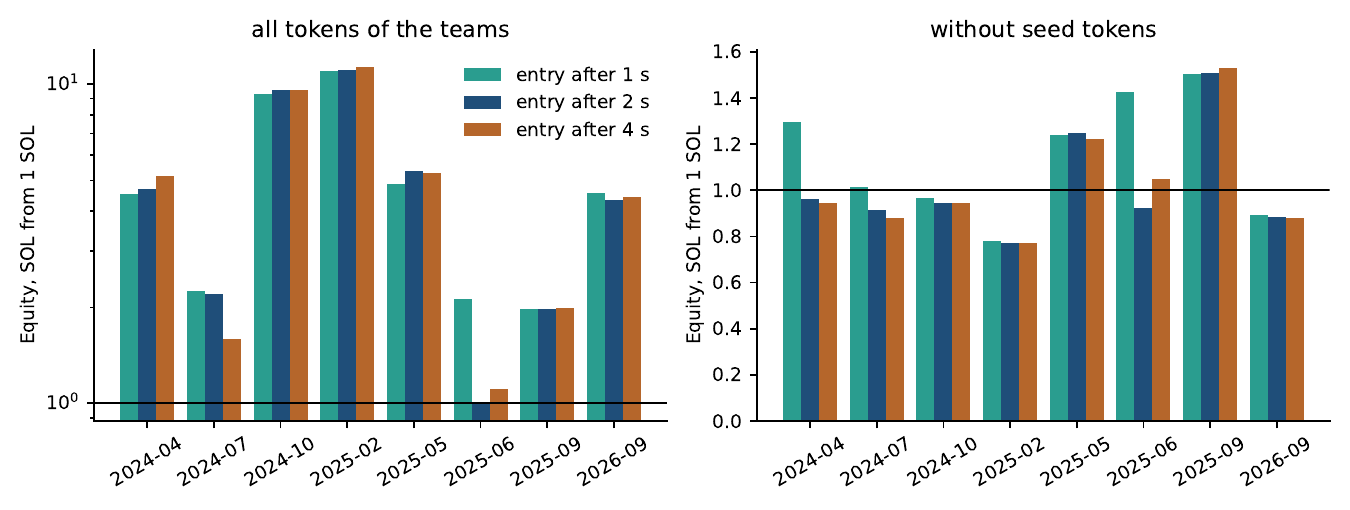}\caption{Equity of the main account by window for entry after 1, 2 and 4 seconds, all tokens (left, logarithmic axis) and without seed tokens (right). The line is one SOL.}\label{fig:strat_delay}\end{figure}

Entering after one second instead of four gives a geometric mean equity of 4.20 SOL (all tokens) and 1.11 (without seed tokens), against 3.82 and 1.00 at four seconds and 3.87 and 1.00 at two seconds (Figure~\ref{fig:strat_delay}). One second is better than four in 3 windows for all tokens (2024-07, 2025-06, 2026-09) and worse in 5 (2024-04, 2024-10, 2025-02, 2025-05, 2025-09), and in 7 windows without seed tokens; the gain is not systematic. The advantage over the random control with the same threshold search is kept at all three delays (at most 25 of 1,000 random sets reach the team account in 8, 7 and 8 of 8 windows for one, two and four seconds). The reserve archive has one state per second, so one second means the next recorded state after creation, not an instantaneous price.

\subsection{Position size}
\begin{figure}[htbp]\centering\includegraphics[width=.9\linewidth]{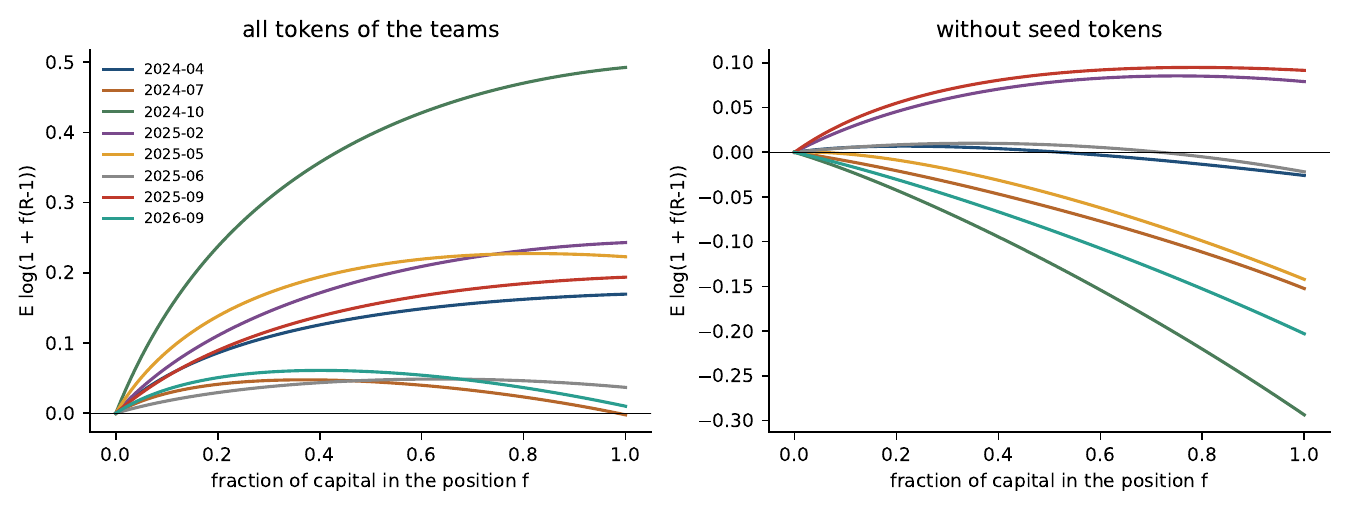}\caption{Log growth $E\ln(1+f(R-1))$ of the standard position by fraction of capital $f$, for each window, all tokens (left) and without seed tokens (right).}\label{fig:strat_kelly}\end{figure}

If a position is multiplied by $R$ and a fraction $f$ of capital is invested, the log growth is $g(f)=E[\ln(1+f(R-1))]$ and the Kelly fraction~\cite{kelly} maximises it. For the standard position of the team tokens the median of $R$ is between 0.76 and 0.96, the mean between 1.22 and 2.83, and between 4\% and 39\% of the positions exceed twice the entry value (Figure~\ref{fig:strat_kelly}). With all tokens the maximiser is at the boundary or high (1.00, 0.37, 1.00, 1.00, 0.82, 0.63, 1.00, 0.40 by window), because a few tokens carry the sample; without seed tokens it is 0.21, 0.00, 0.00, 0.75, 0.02, 0.35, 0.78, 0.00 and $g$ is nearly flat, so the sample gives no support for a size at all. The size that the market allows is limited by price impact: an own purchase of $a$ SOL moves the price of a pool with reserve $R$ by about $2a/(30+R)$, that is 2.9\% for 0.5 SOL at a reserve of 5 SOL and 0.29\% for 0.05 SOL.

\begin{figure}[htbp]\centering\includegraphics[width=\linewidth]{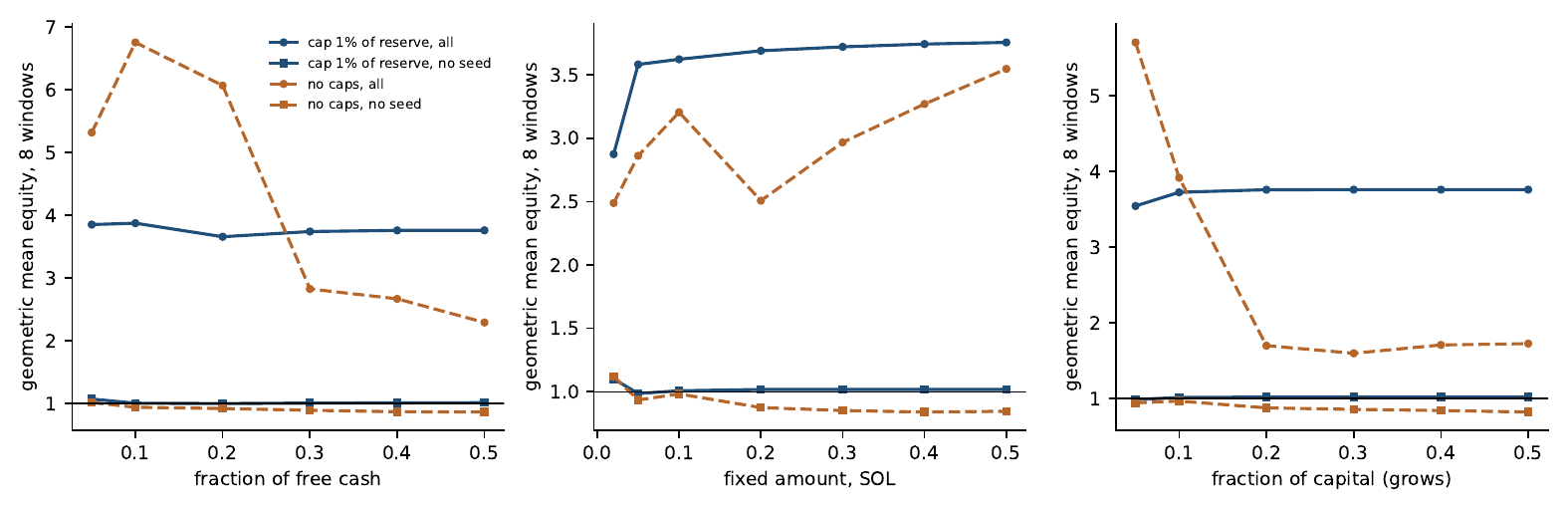}\caption{Geometric mean over the eight windows of equity for a fraction of free cash, a fixed amount and a fraction of the growing capital, with a cap of 1\% of the entry reserve and without caps; all tokens (circles) and without seed tokens (squares).}\label{fig:strat_sizing}\end{figure}

Table~\ref{tab:strat_sizing} and Figure~\ref{fig:strat_sizing} compare the schemes: a fraction of the free cash (the base rule), a fixed amount, a fixed fraction of the initial deposit (a fixed amount of 0.2 to 0.5 SOL) and a fraction of the growing capital, each with the cap of one percent of the entry reserve, without caps, and with only the 0.1 SOL cap. Fixed amounts and fractions of the initial deposit give 3.72 (0.3 SOL) with the reserve cap, a fraction of growing capital 3.76, against 3.82 for the base rule; without seed tokens all of them are between 0.84 and 1.12. Without the reserve cap the results become a bet on a few tokens: a fixed amount of 0.2 to 0.5 SOL gives 24 to 58 SOL in 2024-10 but ends below one SOL in 2024-04, 2024-07, 2025-05, 2025-09. Growing the position with capital does not help, because the capital grows little and is almost all locked in open positions. Realised drawdown is close to zero, since almost nothing is closed, so the lowest window equity is the better measure of risk (0.76 to 0.83 SOL for fixed amounts of 0.2 to 0.5 SOL without caps).

\begin{table}[htbp]\centering\footnotesize\setlength{\tabcolsep}{3pt}
\caption{Position sizing, pooled over the 8 windows (baseline team thresholds, entry after 4 s). The full grid (fractions of cash and of capital 5 to 50\%, fixed amounts 0.02 to 0.5 SOL, four cap settings) is in the supplementary tables.}\label{tab:strat_sizing}
\begin{tabular}{p{5.2cm}rrrrrr}\toprule
Scheme & All: geo.\ mean & Worst & $>1$ & No seed: geo.\ mean & $>1$ & 2024-10 equity \\\midrule
Cash fraction 10\% (v8; caps 0.1 SOL, 1\% of reserve) & 3.82 & 1.10 & 8/8 & 1.00 & 3/8 & 9.6 \\
Cash fraction 20\%, no caps & 6.07 & 0.85 & 6/8 & 0.92 & 2/8 & 258.4 \\
Fixed 0.05 SOL, cap 1\% of reserve & 3.58 & 1.09 & 8/8 & 0.98 & 3/8 & 6.7 \\
Fixed 0.1 SOL, no caps & 3.21 & 0.85 & 5/8 & 0.98 & 1/8 & 12.7 \\
Fixed 0.3 SOL (30\% of the initial deposit), cap 1\% of reserve & 3.72 & 1.25 & 8/8 & 1.02 & 4/8 & 9.7 \\
Fixed 0.5 SOL, no caps & 3.55 & 0.79 & 4/8 & 0.84 & 1/8 & 57.5 \\
Fraction 20\% of growing capital, cap 1\% of reserve & 3.76 & 1.25 & 8/8 & 1.02 & 4/8 & 9.7 \\
Fraction 5\% of growing capital, no caps & 5.71 & 0.93 & 6/8 & 0.94 & 2/8 & 151.4 \\
\bottomrule\end{tabular}\end{table}

\subsection{Capital locked in open positions}
\begin{figure}[htbp]\centering\includegraphics[width=\linewidth]{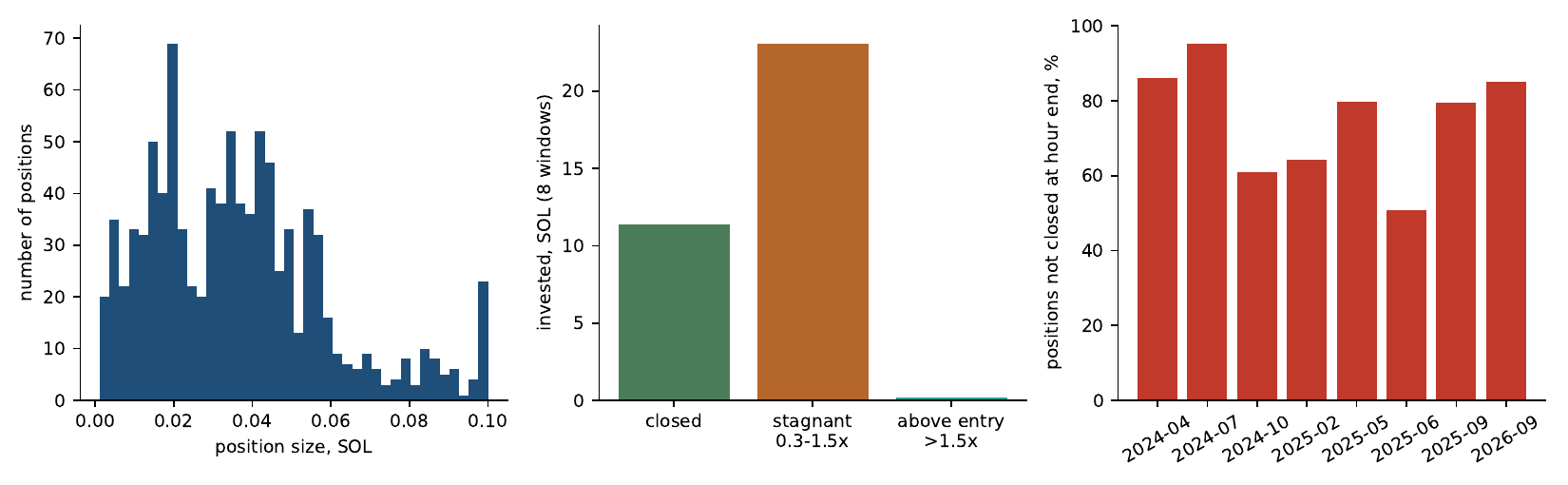}\caption{Positions of the main accounts in the eight windows: size distribution (left), SOL invested by outcome (middle) and the share of positions that are not closed at the end of the hour in every window (right).}\label{fig:strat_stuck}\end{figure}

Of the 947 positions that the main accounts open in the eight windows, 251 (27\%) are closed at the team threshold; 73\% of the positions and 67\% of the invested SOL stay open (Table~\ref{tab:strat_locked}, Figure~\ref{fig:strat_stuck}). Almost all are in a stagnant state: 688 positions are worth 0.3 to 1.5 of the entry value, 8 are above 1.5 but below the threshold, and none has fallen below 0.3, because the price is proportional to $(30+R)^2$ and a loss of the real reserve of a few SOL reduces the price only to about 0.7 of the entry price. A position that has lost its reserve is therefore still worth about 0.6 to 0.9 of what was paid, depending on the entry reserve, and that remainder is locked.

The positions are small because the cap of one percent of the entry reserve binds in 82\% of them (median size 0.034 SOL), the cash fraction in 16\% and the 0.1 SOL cap in 2.1\%. Positions below 0.01 SOL are 11\% of the number but only 0.64 of 34.7 SOL invested. The account runs out of free cash early: the first refusal because of the cash stop falls between 3.1 and 46.2 hours after the first token without seed tokens, and 9 of the 16 accounts (all tokens and without seed tokens in eight windows) reach the stop before the end of the window; after that, between 6 and 90\% of the window is spent without trading. An Erlang-type blocking model does not apply, because the open positions are not released, so the account holds roughly the free cash divided by the mean cost of a position, about 25 positions.

\begin{table}[htbp]\centering\footnotesize\setlength{\tabcolsep}{3pt}
\caption{Opened positions of the main account in the 8 windows (947 positions): outcome at the end of the first hour, SOL invested (including fees) and median position size. Shares are of the total and are rounded.}\label{tab:strat_locked}
\begin{tabular}{p{5cm}rrrrr}\toprule
Class & Positions & Share & SOL invested & Share of SOL & Median size, SOL \\\midrule
Closed at the team threshold & 251 & 27\% & 11.4 & 33\% & 0.038 \\
Open, 0.3 to 1.5 times the entry value & 688 & 73\% & 23.1 & 67\% & 0.031 \\
Open, above 1.5 times, below the threshold & 8 & 1\% & 0.2 & 1\% & 0.020 \\
\bottomrule\end{tabular}\end{table}

Selling all positions that are still open at the last admissible state of the first hour raises the geometric mean equity from 3.82 to 5.65 (without seed tokens 1.62 to 1.86) and is not worse in any of the eight windows (2024-07: 1.59 to 5.26; 2025-06: 1.10 to 5.75). It does not create value: the position was already marked at that state. It converts the mark into cash and allows later entries. Without seed tokens the sale makes a loss visible in 2026-09 (1.09 to 0.89). The assumption is that a pool can be sold into at its recorded reserve at the end of the hour, which fails if the creator has removed the liquidity. Shorter holding limits (60 minutes, 30, 10 and 5 minutes with the same sale) do not improve on selling at the hour end: 3.95, 5.47, 4.52 and 3.98.

\subsection{Exit at a collapse}
\begin{figure}[htbp]\centering\includegraphics[width=\linewidth]{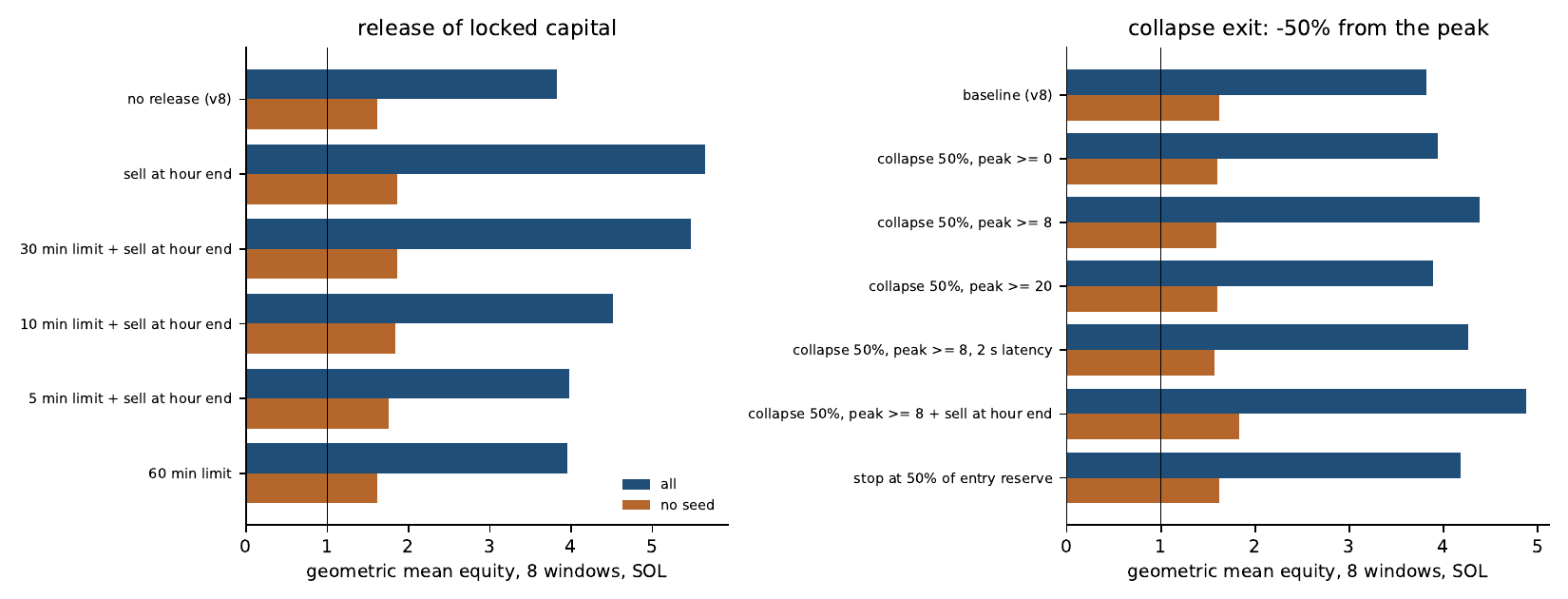}\caption{Geometric mean over the eight windows of equity for rules that free locked capital (left) and for exits at a collapse (right); all tokens and without seed tokens. The team thresholds are chosen again for every rule.}\label{fig:strat_collapse}\end{figure}

A collapse is a fall of the closing reserve by half from its running peak since launch, with a minimum peak so that early noise does not trigger it; the exit is the first recorded state at least a latency after the trigger. The reserve is used rather than the price because the price moves less: a fall of the reserve by half from a peak of 40 SOL means 0.51 of the peak price, but only 0.80 from a peak of 8 SOL. Without such a rule the base account exits only at its upward threshold, so a token that has lost most of its reserve remains open until the end of the window. Table~\ref{tab:strat_rules} and Figure~\ref{fig:strat_collapse} give the results. A collapse exit with no minimum peak gives a geometric mean of 3.94 (without seed 1.60) against 3.82; the worst window rises from 1.10 to 1.33, in 2025-06 from 1.10 to 4.63. In several windows the rule lowers equity (2025-05: 5.26 to 2.31; 2025-02: 11.32 to 8.67; 2024-04: 5.14 to 3.65), because some tokens recover after a fall by half and some falls are retreats after a large gain. A latency of two or five seconds changes the geometric mean by 0.03 to 0.15, so a delay of a few seconds does not matter for a large collapse. The simplest rule, a stop at half of the entry reserve, gives 4.18 (1.61 without seed), and selling at the hour end alone gives 5.65; combining them adds nothing (4.88), since all of them gain from freeing cash. Against the random control with the same threshold search every rule keeps the advantage in 7 to 8 of 8 windows.

\subsection{Collapse by team}
\begin{figure}[htbp]\centering\includegraphics[width=\linewidth]{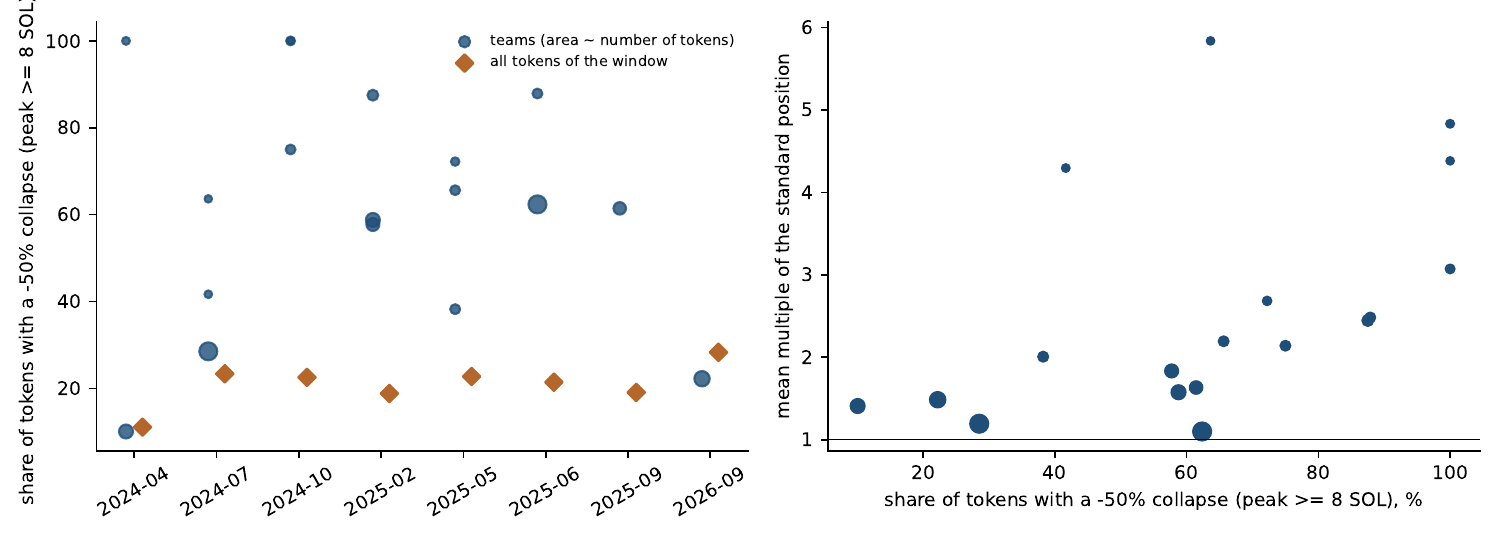}\caption{Share of tokens with a collapse of 50\% of the reserve from a peak of at least 8 SOL: teams by window against all tokens of the window (left), and the mean multiple of the standard position against that share for each team (right).}\label{fig:strat_teams}\end{figure}

Among all eligible tokens of a window 11 to 28\% have a collapse by this definition, 5 to 24\% a fast one (at most a minute from the peak). The tokens of the teams collapse more often, between 18\% (2024-04) and 88\% (2024-10) by window, and the spread between teams is wider (from 10\% to 100\%) (Table~\ref{tab:strat_teams}, Figure~\ref{fig:strat_teams}). This is partly selection: the tokens that the team pushes to a high reserve are exactly those that can fall by half from a peak of 8 SOL. The tokens with a collapse do not lose on average: their mean multiple under the team threshold is above that of the tokens without a collapse in 8 of 8 windows (2024-04: 4.55 against 1.09; 2025-02: 2.37 against 0.87), because the collapse follows a rise and the threshold has often been reached. A share between 8\% and 78\% of the team tokens collapses before the team threshold is reached; for them the collapse exit replaces waiting. A reserve collapse is not evidence of a rug pull, which would require the sales of the creator; those events were not used here.

\begin{table}[htbp]\centering\footnotesize\setlength{\tabcolsep}{3pt}
\caption{Collapse of the tokens of the teams (all teams of a window together). A collapse is a fall of the closing reserve by 50\% from its running peak of at least 8 SOL after entry. Shares of tokens: collapse among the team tokens and, for comparison, among all eligible tokens of the window; fast collapse (at most 60 s from the peak to the trigger); collapse before the team threshold is reached. Last three columns: mean multiple of the standard position under the team threshold, for all team tokens, for those with a collapse and without.}\label{tab:strat_teams}
\begin{tabular}{lrrrrrrrr}\toprule
Window & Tokens & Collapse & Window & Fast & Before thr. & Mult. all & Collapsed & Not coll. \\\midrule
2024-04 & 142 & 18\% & 11\% & 8\% & 10\% & 1.72 & 4.55 & 1.09 \\
2024-07 & 377 & 30\% & 23\% & 17\% & 27\% & 1.43 & 1.97 & 1.20 \\
2024-10 & 64 & 88\% & 23\% & 61\% & 55\% & 2.83 & 3.13 & 0.71 \\
2025-02 & 276 & 63\% & 19\% & 56\% & 34\% & 1.82 & 2.37 & 0.87 \\
2025-05 & 84 & 56\% & 23\% & 50\% & 44\% & 2.22 & 2.25 & 2.18 \\
2025-06 & 389 & 65\% & 21\% & 61\% & 21\% & 1.22 & 1.43 & 0.83 \\
2025-09 & 83 & 61\% & 19\% & 57\% & 53\% & 1.63 & 2.06 & 0.96 \\
2026-09 & 189 & 22\% & 28\% & 21\% & 19\% & 1.49 & 1.75 & 1.41 \\
\bottomrule\end{tabular}\end{table}

\paragraph{Limits.} All thresholds are chosen in hindsight, the rules were selected after the results were seen, the sale at the hour end and the collapse exit assume execution at the recorded reserve, the study was not repeated on windows without a team, and no rule was tested in the prospective setting of Section~\ref{sec:prospective}.
\FloatBarrier

\section{Worked examples and additional tables}
\subsection{The 85-SOL boundary on one token}\label{sec:app_boundary}

\begin{figure}[htbp]\centering\includegraphics[width=.85\linewidth]{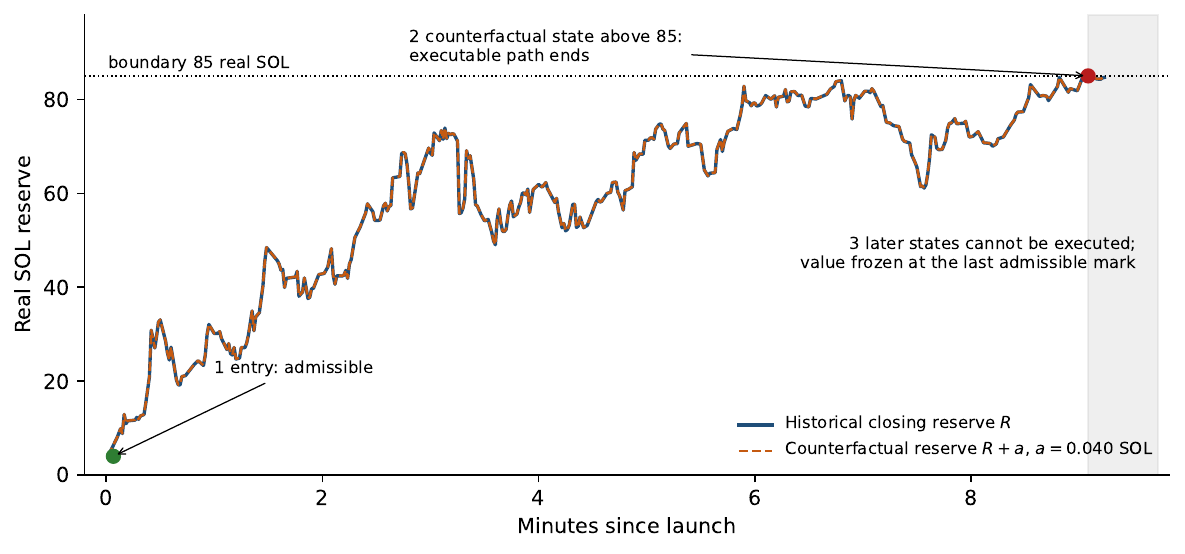}\caption{Historical and counterfactual reserve of one real token and the three states of a position at the 85-SOL boundary.}\label{fig:v8_boundary}\end{figure}

\paragraph{The boundary in one example.} Figure~\ref{fig:v8_boundary} follows one real token. The position is entered four seconds after launch at an entry reserve of 4.0 real SOL with a principal of 0.040 SOL (state 1, admissible). The historical reserve reaches 84.989 SOL after 9.1 minutes; with the follower's deposit added, the counterfactual reserve exceeds 85 SOL at that moment (state 2), so the executable path of the position ends: its value is frozen at the last admissible mark, later states are not used even if the reserve fell back, and no cash is released (state 3). A frozen position is a censored observation, not a sale. The limit of 85 SOL is a conservative boundary of our model domain, not a claim about the protocol: the documented completion state lies slightly above 85 SOL, so crossing near completion removes opportunities rather than proving a data error.

\subsection{Equity, cash and marks in numbers}\label{sec:app_equity}
\paragraph{A numerical example of equity, cash and marks.} Consider two positions of the example team of Appendix~\ref{sec:example}. The first token is bought four seconds after launch at an entry reserve of 9.58 real SOL for 0.096 SOL (plus a one percent fee). It reaches the team's threshold after 6.2 minutes at a reserve of 64.09 SOL, the model sells and releases 0.533 SOL of cash. The second token is bought at a reserve of 8.69 SOL for 0.087 SOL, never reaches the threshold within its hour and ends at a reserve of 0.00 SOL; the model marks it at 0.052 SOL, -41\% relative to the entry debit. Both appear in equity, but only the first has produced cash. The second is locked: it cannot finance another entry and would have to be sold into a pool with little liquidity left. A window whose equity is mostly marks therefore says little about money that could have been spent.

\begin{table}[htbp]\centering\footnotesize\setlength{\tabcolsep}{2.5pt}
\caption{Every trade of the example team (2024-10, team 2, threshold 64 real SOL). Launch time is UTC; entry four seconds after launch; size is the principal in SOL; hold is minutes from entry to the exit or to the end of the first hour; reserves are real SOL at entry and at exit (or at the last admissible state); proceeds are SOL released at the exit, or the model mark for an open position; return is relative to the entry debit including the 1\% fee.}\label{tab:v8_example}
\begin{tabular}{rrrrrrlrr}\toprule
No. & Launch & Entry res. & Size & Hold & Exit res. & Outcome & Proceeds & Return \\\midrule
1 & 01:16:38 & 9.58 & 0.096 & 6.2 & 64.09 & threshold reached & 0.533 & +450\% \\
2 & 01:27:17 & 8.00 & 0.080 & 4.7 & 66.93 & threshold reached & 0.512 & +534\% \\
3 & 01:38:37 & 8.69 & 0.087 & 59.9 & 0.00 & open (marked) & 0.052 & -41\% \\
4 & 01:40:15 & 8.47 & 0.085 & 59.9 & 0.00 & open (marked) & 0.051 & -40\% \\
5 & 01:42:04 & 8.30 & 0.083 & 2.4 & 66.36 & threshold reached & 0.517 & +517\% \\
6 & 01:50:45 & 8.40 & 0.084 & 59.9 & 0.01 & open (marked) & 0.051 & -40\% \\
7 & 02:02:21 & 9.89 & 0.099 & 4.0 & 65.01 & threshold reached & 0.552 & +453\% \\
8 & 02:13:48 & 8.38 & 0.084 & 3.9 & 64.19 & threshold reached & 0.497 & +487\% \\
9 & 02:23:26 & 8.52 & 0.085 & 3.8 & 64.16 & threshold reached & 0.501 & +482\% \\
10 & 02:32:26 & 9.67 & 0.097 & 4.8 & 64.26 & threshold reached & 0.537 & +450\% \\
11 & 02:46:38 & 10.10 & 0.100 & 3.8 & 64.71 & threshold reached & 0.549 & +443\% \\
\bottomrule\end{tabular}\end{table}
\FloatBarrier

\subsection{Why tokens are not bought, and equity by component}\label{sec:app_tables}
\begin{table}[htbp]\centering\footnotesize\setlength{\tabcolsep}{2.5pt}
\caption{Why selected tokens are not bought. Starting from the union of the selected teams: tokens without an admissible entry state (no fresh reserve at the delay, or a reserve of zero), tokens met while free cash was below the 0.1 SOL stop, other rejections (order below the minimum size), positions opened, closed at their threshold and still open. Causes are exclusive and tested in this order.}\label{tab:v8_funnel}
\begin{tabular}{lrrrrrrr}\toprule
Window & Union & No entry state & Cash below stop & Other & Opened & Closed & Open \\\midrule
2024-04 & 142 & 1 & 0 & 4 & 137 & 19 & 118 \\
2024-07 & 379 & 2 & 313 & 0 & 64 & 3 & 61 \\
2024-10 & 64 & 0 & 0 & 0 & 64 & 25 & 39 \\
2025-02 & 276 & 0 & 0 & 11 & 265 & 95 & 170 \\
2025-05 & 84 & 0 & 0 & 0 & 84 & 17 & 67 \\
2025-06 & 389 & 0 & 279 & 0 & 110 & 54 & 56 \\
2025-09 & 83 & 0 & 0 & 0 & 83 & 17 & 66 \\
2026-09 & 189 & 0 & 0 & 49 & 140 & 21 & 119 \\
\bottomrule\end{tabular}\end{table}
\begin{table}[htbp]\centering\footnotesize\setlength{\tabcolsep}{2.5pt}
\caption{Equity, free cash and marks. Equity $=$ cash $+$ marks. The realised result of closed trades is cash earned above cost; marks are the model value of positions that never reached their exit threshold, relative to their cost. Marks cannot finance new entries and are not proceeds.}\label{tab:v8_decomp}
\begin{tabular}{lrrrrrrr}\toprule
Window & Equity & Cash & Marks & Share & Closed net & Open cost & Marks/cost \\\midrule
2024-04 & 5.14 & 3.22 & 1.92 & 37\% & 4.27 & 2.05 & 0.94 \\
2024-07 & 1.59 & 0.10 & 1.49 & 94\% & 0.73 & 1.63 & 0.91 \\
2024-10 & 9.61 & 8.00 & 1.61 & 17\% & 9.29 & 2.29 & 0.70 \\
2025-02 & 11.32 & 6.09 & 5.23 & 46\% & 12.26 & 7.17 & 0.73 \\
2025-05 & 5.26 & 3.35 & 1.91 & 36\% & 4.87 & 2.52 & 0.76 \\
2025-06 & 1.10 & 0.10 & 1.01 & 91\% & 0.57 & 1.48 & 0.68 \\
2025-09 & 1.99 & 0.83 & 1.16 & 58\% & 1.21 & 1.38 & 0.84 \\
2026-09 & 4.43 & 1.20 & 3.24 & 73\% & 4.97 & 4.77 & 0.68 \\
\bottomrule\end{tabular}\end{table}
\FloatBarrier

\section{Additional window-level results}
\par
\begingroup\small\setlength{\tabcolsep}{4pt}
\begin{longtable}{lllr}
\caption{Recorded observation bounds. Filenames and month labels are identifiers; these are the actual source bounds. The windows are purposive samples, not whole-month censuses.}\\
\toprule
Window & Start UTC & End UTC & Hours\\\midrule\endfirsthead
\toprule
Window & Start UTC & End UTC & Hours\\\midrule\endhead
\bottomrule\endfoot
2024-04 & 2024-04-10 19:57:42 & 2024-04-11 19:38:40 & 23.68\\
2024-07 & 2024-07-10 19:59:09 & 2024-07-12 03:25:04 & 31.43\\
2024-10 & 2024-10-10 19:59:59 & 2024-10-12 05:20:28 & 33.34\\
2025-02 & 2025-02-10 19:59:59 & 2025-02-12 02:58:55 & 30.98\\
2025-03 & 2025-03-10 19:59:59 & 2025-03-11 19:40:31 & 23.68\\
2025-05 & 2025-05-10 19:59:59 & 2025-05-13 03:24:12 & 55.40\\
2025-06 & 2025-06-21 19:59:59 & 2025-06-25 03:16:26 & 79.27\\
2025-09 & 2025-09-12 19:59:59 & 2025-09-15 19:18:29 & 71.31\\
2025-12 & 2025-12-12 19:59:59 & 2025-12-15 19:36:56 & 71.62\\
2026-09 & 2026-09-27 15:00:00 & 2026-09-29 15:00:00 & 48.00\\
\end{longtable}\endgroup

\par
\begingroup\small\setlength{\tabcolsep}{4pt}
\begin{longtable}{lrrrrr}
\caption{Cohort attrition. Full hour means the archive extends past the token's first hour, not that every trade was observed. Mayhem or unknown-mode exclusion and absence of an admitted state in the first minute are disjoint within the full-hour cohort.}\\
\toprule
Window & Launches & Full hour & Mode excluded & No early state & Eligible\\\midrule\endfirsthead
\toprule
Window & Launches & Full hour & Mode excluded & No early state & Eligible\\\midrule\endhead
\bottomrule\endfoot
2024-04 & 11,916 & 11,527 & 0 & 0 & 11,527\\
2024-07 & 14,282 & 13,920 & 0 & 9 & 13,911\\
2024-10 & 22,398 & 21,955 & 0 & 27 & 21,928\\
2025-02 & 73,394 & 71,262 & 0 & 163 & 71,099\\
2025-03 & 25,856 & 24,013 & 0 & 3 & 24,010\\
2025-05 & 57,939 & 57,327 & 0 & 197 & 57,130\\
2025-06 & 84,821 & 83,907 & 0 & 450 & 83,457\\
2025-09 & 81,233 & 78,940 & 0 & 1,280 & 77,660\\
2025-12 & 52,377 & 51,422 & 5,923 & 2,048 & 43,451\\
2026-09 & 83,742 & 81,836 & 24,658 & 11,582 & 45,596\\
\end{longtable}\endgroup

\par
\begingroup\small\setlength{\tabcolsep}{4pt}
\begin{longtable}{lrrrrr}
\caption{Observed real-reserve maxima. Implied spot moves are reference-curve quantities, not recorded prices or account returns.}\\
\toprule
Window & Tokens & Median max & Ninetieth pct. & Maximum & Implied twofold\\\midrule\endfirsthead
\toprule
Window & Tokens & Median max & Ninetieth pct. & Maximum & Implied twofold\\\midrule\endhead
\bottomrule\endfoot
2024-04 & 11,527 & 3.023 & 9.594 & 85.000 & 5.75\%\\
2024-07 & 13,911 & 3.995 & 23.133 & 84.999 & 9.42\%\\
2024-10 & 21,928 & 3.678 & 20.903 & 85.000 & 9.58\%\\
2025-02 & 71,099 & 2.875 & 19.184 & 85.000 & 7.78\%\\
2025-03 & 24,010 & 2.970 & 15.447 & 84.976 & 6.61\%\\
2025-05 & 57,130 & 3.752 & 16.301 & 84.998 & 6.60\%\\
2025-06 & 83,457 & 3.493 & 15.105 & 84.993 & 6.10\%\\
2025-09 & 77,660 & 2.953 & 14.948 & 84.997 & 6.93\%\\
2025-12 & 43,451 & 2.362 & 15.400 & 84.999 & 7.14\%\\
2026-09 & 45,596 & 2.735 & 21.311 & 84.999 & 8.73\%\\
\end{longtable}\endgroup

\par
\begingroup\small\setlength{\tabcolsep}{4pt}
\begin{longtable}{lrrrrr}
\caption{Illustrative observed reserve reach. Non-reach before censoring is not certified later failure.}\\
\toprule
Window & Threshold & Tokens & Reached & Fraction & Censored\\\midrule\endfirsthead
\toprule
Window & Threshold & Tokens & Reached & Fraction & Censored\\\midrule\endhead
\bottomrule\endfoot
2024-04 & 40.000 & 11,527 & 312 & 2.71\% & 59\\
2024-04 & 80.000 & 11,527 & 89 & 0.77\% & 59\\
2024-07 & 40.000 & 13,911 & 688 & 4.95\% & 166\\
2024-07 & 80.000 & 13,911 & 153 & 1.10\% & 166\\
2024-10 & 40.000 & 21,928 & 936 & 4.27\% & 167\\
2024-10 & 80.000 & 21,928 & 124 & 0.57\% & 167\\
2025-02 & 40.000 & 71,099 & 3,266 & 4.59\% & 681\\
2025-02 & 80.000 & 71,099 & 269 & 0.38\% & 681\\
2025-03 & 40.000 & 24,010 & 693 & 2.89\% & 123\\
2025-03 & 80.000 & 24,010 & 82 & 0.34\% & 123\\
2025-05 & 40.000 & 57,130 & 1,473 & 2.58\% & 228\\
2025-05 & 80.000 & 57,130 & 227 & 0.40\% & 228\\
2025-06 & 40.000 & 83,457 & 1,888 & 2.26\% & 394\\
2025-06 & 80.000 & 83,457 & 306 & 0.37\% & 394\\
2025-09 & 40.000 & 77,660 & 1,976 & 2.54\% & 457\\
2025-09 & 80.000 & 77,660 & 318 & 0.41\% & 457\\
2025-12 & 40.000 & 43,451 & 1,086 & 2.50\% & 176\\
2025-12 & 80.000 & 43,451 & 100 & 0.23\% & 176\\
2026-09 & 40.000 & 45,596 & 1,703 & 3.73\% & 486\\
2026-09 & 80.000 & 45,596 & 447 & 0.98\% & 486\\
\end{longtable}\endgroup

\par
\begingroup\small\setlength{\tabcolsep}{4pt}
\begin{longtable}{lrrrrr}
\caption{Raw audit after unit normalization. All-launch censor counts differ from entry-eligible counts. Rent inference uses only unique known initial pre-balance classes.}\\
\toprule
Window & Raw rows & Above cap & Censored & Non-Euler & Rent inferred\\\midrule\endfirsthead
\toprule
Window & Raw rows & Above cap & Censored & Non-Euler & Rent inferred\\\midrule\endhead
\bottomrule\endfoot
2024-04 & 226,241 & 146 & 59 & 0 & 0\\
2024-07 & 965,762 & 355 & 173 & 0 & 0\\
2024-10 & 1,686,113 & 393 & 195 & 148 & 0\\
2025-02 & 4,315,437 & 1,820 & 870 & 331 & 0\\
2025-03 & 2,087,992 & 293 & 136 & 172 & 0\\
2025-05 & 4,053,747 & 512 & 247 & 698 & 0\\
2025-06 & 6,139,186 & 1,063 & 513 & 1,134 & 0\\
2025-09 & 4,741,920 & 1,115 & 520 & 1,231 & 0\\
2025-12 & 3,301,915 & 738 & 246 & 1,998 & 0\\
2026-09 & 7,497,852 & 1,329 & 1,220 & 0 & 0\\
\end{longtable}\endgroup

\par
\begingroup\small\setlength{\tabcolsep}{4pt}
\begin{longtable}{llrrrr}
\caption{Admissible-state retention sensitivities with common endpoints. Equity uses the common reserve exit, not the team thresholds of the main account.}\\
\toprule
Window & Filter & Eligible & Selected & Equity & Cash\\\midrule\endfirsthead
\toprule
Window & Filter & Eligible & Selected & Equity & Cash\\\midrule\endhead
\bottomrule\endfoot
2024-04 & primary & 11,527 & 2 & 2.970 & 1.159\\
2024-04 & reserve\_changes & 11,509 & 2 & 2.970 & 1.159\\
2024-04 & sol\_checked & 11,455 & 2 & 3.023 & 1.233\\
2024-07 & primary & 13,911 & 3 & 1.371 & 0.096\\
2024-07 & reserve\_changes & 13,850 & 3 & 1.371 & 0.096\\
2024-07 & sol\_checked & 13,845 & 3 & 1.378 & 0.096\\
2024-10 & primary & 21,928 & 3 & 5.380 & 3.893\\
2024-10 & reserve\_changes & 21,918 & 3 & 5.380 & 3.893\\
2024-10 & sol\_checked & 21,917 & 3 & 5.380 & 3.893\\
2025-02 & primary & 71,099 & 3 & 3.592 & 0.093\\
2025-02 & reserve\_changes & 68,190 & 3 & 3.591 & 0.093\\
2025-02 & sol\_checked & 68,019 & 3 & 3.836 & 0.097\\
2025-03 & primary & 24,010 & 0 & n.a. & n.a.\\
2025-03 & reserve\_changes & 23,314 & 0 & n.a. & n.a.\\
2025-03 & sol\_checked & 23,364 & 0 & n.a. & n.a.\\
2025-05 & primary & 57,130 & 3 & 3.258 & 1.564\\
2025-05 & reserve\_changes & 55,837 & 3 & 3.258 & 1.564\\
2025-05 & sol\_checked & 56,259 & 3 & 3.280 & 1.583\\
2025-06 & primary & 83,457 & 3 & 0.793 & 0.095\\
2025-06 & reserve\_changes & 80,777 & 3 & 0.793 & 0.095\\
2025-06 & sol\_checked & 82,967 & 3 & 0.941 & 0.094\\
2025-09 & primary & 77,660 & 1 & 2.007 & 0.830\\
2025-09 & reserve\_changes & 72,995 & 1 & 2.007 & 0.829\\
2025-09 & sol\_checked & 73,134 & 1 & 2.005 & 0.822\\
2025-12 & primary & 43,451 & 0 & n.a. & n.a.\\
2025-12 & reserve\_changes & 38,972 & 0 & n.a. & n.a.\\
2025-12 & sol\_checked & 38,751 & 0 & n.a. & n.a.\\
2026-09 & primary & 45,596 & 1 & 2.625 & 0.100\\
2026-09 & reserve\_changes & 40,993 & 1 & 2.625 & 0.100\\
2026-09 & sol\_checked & 42,623 & 1 & 2.263 & 0.098\\
\end{longtable}\endgroup

\par
\begingroup\small\setlength{\tabcolsep}{4pt}
\begin{longtable}{lrrrrr}
\caption{Seed-free five-minute/stop entry-delay sensitivity, still using retrospective group identities.}\\
\toprule
Window & Delay (s) & Equity & Cash & Unresolved & Random\\\midrule\endfirsthead
\toprule
Window & Delay (s) & Equity & Cash & Unresolved & Random\\\midrule\endhead
\bottomrule\endfoot
2024-04 & 4 & 1.046 & 0.429 & 34 & 0.848\\
2024-04 & 8 & 1.035 & 0.393 & 34 & 0.824\\
2024-04 & 15 & 0.977 & 0.334 & 33 & 0.807\\
2024-04 & 30 & 1.022 & 0.313 & 33 & 0.734\\
2024-07 & 4 & 1.056 & 0.099 & 57 & 0.456\\
2024-07 & 8 & 0.930 & 0.097 & 48 & 0.457\\
2024-07 & 15 & 0.905 & 0.096 & 49 & 0.456\\
2024-07 & 30 & 0.900 & 0.094 & 47 & 0.473\\
2024-10 & 4 & 1.614 & 1.614 & 0 & 0.851\\
2024-10 & 8 & 1.597 & 1.597 & 0 & 0.851\\
2024-10 & 15 & 1.599 & 1.599 & 0 & 0.869\\
2024-10 & 30 & 1.545 & 1.545 & 0 & 0.884\\
2025-02 & 4 & 1.035 & 0.927 & 4 & 0.406\\
2025-02 & 8 & 0.491 & 0.402 & 5 & 0.404\\
2025-02 & 15 & 0.173 & 0.119 & 5 & 0.430\\
2025-02 & 30 & 0.150 & 0.097 & 6 & 0.482\\
2025-05 & 4 & 0.998 & 0.738 & 11 & 0.664\\
2025-05 & 8 & 1.200 & 0.944 & 11 & 0.698\\
2025-05 & 15 & 1.149 & 0.910 & 11 & 0.743\\
2025-05 & 30 & 0.873 & 0.696 & 11 & 0.803\\
2025-06 & 4 & 0.151 & 0.099 & 12 & 0.241\\
2025-06 & 8 & 0.163 & 0.100 & 17 & 0.248\\
2025-06 & 15 & 0.183 & 0.099 & 26 & 0.277\\
2025-06 & 30 & 0.243 & 0.098 & 50 & 0.352\\
2025-09 & 4 & 1.482 & 1.482 & 0 & 0.750\\
2025-09 & 8 & 1.544 & 1.521 & 4 & 0.777\\
2025-09 & 15 & 1.545 & 1.503 & 7 & 0.803\\
2025-09 & 30 & 1.408 & 1.357 & 9 & 0.840\\
2026-09 & 4 & 0.176 & 0.108 & 8 & 0.380\\
2026-09 & 8 & 0.394 & 0.306 & 12 & 0.430\\
2026-09 & 15 & 0.575 & 0.473 & 12 & 0.477\\
2026-09 & 30 & 0.803 & 0.676 & 13 & 0.564\\
\end{longtable}\endgroup

\par
\begingroup\small\setlength{\tabcolsep}{4pt}
\begin{longtable}{lrrrrr}
\caption{After-seed five-minute/stop at base costs. The full seed rule must be observable, but group identities and membership remain retrospective.}\\
\toprule
Window & Tokens & Equity & Cash & Unresolved & Random\\\midrule\endfirsthead
\toprule
Window & Tokens & Equity & Cash & Unresolved & Random\\\midrule\endhead
\bottomrule\endfoot
2024-04 & 109 & 1.956 & 1.392 & 28 & 0.931\\
2024-07 & 356 & 2.137 & 0.677 & 55 & 0.574\\
2024-10 & 60 & 3.593 & 3.593 & 0 & 0.901\\
2025-02 & 251 & 2.729 & 2.614 & 4 & 0.512\\
2025-05 & 74 & 1.253 & 0.733 & 11 & 0.685\\
2025-06 & 380 & 1.651 & 1.536 & 19 & 0.262\\
2025-09 & 55 & 1.852 & 1.852 & 0 & 0.861\\
2026-09 & 188 & 3.605 & 3.497 & 8 & 0.491\\
\end{longtable}\endgroup

\par
\begingroup\small\setlength{\tabcolsep}{4pt}
\begin{longtable}{llrrrr}
\caption{Standardized per-token policy means and creation-hour block-bootstrap intervals; not reusable-capital account returns.}\\
\toprule
Window & Cohort & Tokens & Mean net & Block low & Block high\\\midrule\endfirsthead
\toprule
Window & Cohort & Tokens & Mean net & Block low & Block high\\\midrule\endhead
\bottomrule\endfoot
2024-04 & leaders & 137 & 40.02\% & 16.26\% & 68.81\%\\
2024-04 & universe & 11,173 & -1.02\% & -2.17\% & 0.18\%\\
2024-07 & leaders & 377 & 26.23\% & 14.27\% & 40.39\%\\
2024-07 & universe & 13,664 & -5.71\% & -6.63\% & -4.78\%\\
2024-10 & leaders & 64 & 94.96\% & 75.76\% & 122.47\%\\
2024-10 & universe & 21,813 & -6.23\% & -7.59\% & -4.78\%\\
2025-02 & leaders & 265 & 29.77\% & 16.35\% & 48.37\%\\
2025-02 & universe & 63,114 & -6.46\% & -7.32\% & -5.57\%\\
2025-03 & leaders & 0 & n.a. & n.a. & n.a.\\
2025-03 & universe & 22,320 & -10.42\% & -10.95\% & -9.88\%\\
2025-05 & leaders & 84 & 68.45\% & 34.21\% & 112.29\%\\
2025-05 & universe & 52,705 & -10.44\% & -11.04\% & -9.80\%\\
2025-06 & leaders & 389 & 7.04\% & -2.79\% & 16.54\%\\
2025-06 & universe & 73,932 & -11.93\% & -12.35\% & -11.43\%\\
2025-09 & leaders & 83 & 63.08\% & 36.49\% & 91.46\%\\
2025-09 & universe & 62,247 & -7.40\% & -7.85\% & -6.94\%\\
2025-12 & leaders & 0 & n.a. & n.a. & n.a.\\
2025-12 & universe & 33,089 & -8.82\% & -9.32\% & -8.32\%\\
2026-09 & leaders & 140 & 55.45\% & 26.69\% & 107.14\%\\
2026-09 & universe & 35,032 & -9.82\% & -10.52\% & -9.07\%\\
\end{longtable}\endgroup

\section{The separately bounded September 2026 sample}\label{sec:sept2026}
The additional sample, collected directly from the Solana blockchain, covers only 27 September 2026 15:00:00 UTC through 29 September 2026 15:00:00 UTC, an exclusive-end, central 48-hour interval fixed before ranking or trading outcomes. Tokens born during its final hour cannot enter the completed-hour cohort. The full collected record and shorter common-coverage alternatives are not additional experimental windows. Its teams, endpoint incidence, controls and midpoint selection are reconstructed solely within these 48 hours, without transferring historical identities. 

This source records observed creation instructions separately from incomplete token placeholders. Placeholders cannot establish launch times or creators. Only explicit non-Mayhem flags and SOL-quoted curve events enter the primary cohort; an unknown mode is excluded separately from confirmed Mayhem. Event real and virtual reserves are NUMERIC lamports, converted to SOL without rent subtraction. Thus the account-balance normalization  applies to the earlier time windows, not to the direct event fields in this sample. Unix-second timestamps are converted to milliseconds. Slot, transaction position and chronological emit-event position determine the per-token event sequence, followed by the same completed-second availability rule. Ten blocks missing from the collected data were recovered and checked against the hashes of both neighbouring blocks; the recovered events supplement the collected data.

The primary funding graph uses actual native System Program transfers, including transfers with seed. WSOL transfers between attributed owner addresses enter a separate graph sensitivity. Account creation, rent allocation and nonce withdrawal do not count as funding edges. Earlier archives lack these explicit transfer-type distinctions, so measurement differences remain a limitation of period comparisons. The common one-percent fee per side remains fixed for comparability. An additional 1.25-percent-per-side scenario reflects the recorded ordinary SOL event fee schedule in the new sample, rather than changing the team ranking retrospectively.

\par
\begingroup\small\setlength{\tabcolsep}{4pt}
\begin{longtable}{lr}
\caption{Source attrition and reconstruction for the fixed September 2026 interval. Unknown mode and confirmed Mayhem are distinct exclusions. Native and WSOL edge rows are not numbers of distinct wallets.}\\
\toprule
Source cohort or check & Count\\\midrule\endfirsthead
\toprule
Source cohort or check & Count\\\midrule\endhead
\bottomrule\endfoot
Observed creations & 83,742\\
Full archive hour & 81,836\\
Confirmed Mayhem full hour & 21,777\\
Unknown mode full hour & 2,881\\
Ordinary full hour & 57,178\\
Ordinary entry eligible & 45,596\\
Ordinary no early state & 11,582\\
Final hour excluded & 1,906\\
Recovered blocks & 10\\
Recovered SOL curve events & 73\\
Native transfer rows & 18,102,752\\
WSOL sensitivity rows & 9,394,654\\
\end{longtable}\endgroup
The primary funding rule yields a large descendant set for the selected group. Its interpretation as a coordinated team is especially uncertain: a shared distributor can connect otherwise unrelated creators. The native-plus-WSOL graph and endpoint-incidence surfaces accompany the primary result. No selected identity crosses from another period.
\par
\begingroup\small\setlength{\tabcolsep}{4pt}
\begin{longtable}{llrrrr}
\caption{September 2026 recorded-fee sensitivity: 1.25 percent on each side, zero additional slippage. Group identities stay fixed under the common one-percent ranking configuration.}\\
\toprule
Cohort & Policy & Equity & Cash & Unresolved & Random\\\midrule\endfirsthead
\toprule
Cohort & Policy & Equity & Cash & Unresolved & Random\\\midrule\endhead
\bottomrule\endfoot
all & exit40 & 2.560 & 0.098 & 104 & 0.783\\
seed-free & exit40 & 0.904 & 0.091 & 30 & 0.735\\
all & hold300s\_stop50 & 3.552 & 3.445 & 8 & 0.458\\
seed-free & hold300s\_stop50 & 0.169 & 0.102 & 8 & 0.373\\
\end{longtable}\endgroup
\par
\begingroup\small\setlength{\tabcolsep}{4pt}
\begin{longtable}{lrrrrr}
\caption{September 2026 graph sensitivity. Adding owner-attributed WSOL edges changes both reachability and endpoint incidence; it does not merely add trades to a fixed team. No qualifying team means no following account. The accounts use the common reserve exit (one threshold for all groups), not the team-specific thresholds of the main account, so the baseline equity (2.625 SOL) differs from the 4.43 SOL of the main account.}\\
\toprule
Funding variant & Candidates & Selected & Union tokens & Equity & Cash\\\midrule\endfirsthead
\toprule
Funding variant & Candidates & Selected & Union tokens & Equity & Cash\\\midrule\endhead
\bottomrule\endfoot
ex1000\_sol0 & 123 & 1 & 189 & 2.625 & 0.100\\
ex1000\_sol0\_wsol & 111 & 0 & 0 & n.a. & n.a.\\
ex2000\_sol0 & 115 & 1 & 189 & 2.625 & 0.100\\
ex500\_sol0 & 127 & 3 & 597 & 1.186 & 0.095\\
\end{longtable}\endgroup

\end{document}